\documentclass[preprint,12pt]{elsarticle} 

\usepackage{lineno}
\modulolinenumbers[5]
\usepackage{color}
\usepackage[usenames,dvipsnames]{xcolor}
\usepackage{fullpage}
\usepackage{amsmath}
\usepackage{amssymb}
\usepackage{mathrsfs}
\usepackage{newfloat}
\usepackage{graphicx}
\usepackage{stmaryrd}
\usepackage{subfig}
\usepackage{epstopdf}
\usepackage{multirow}
\usepackage{stackengine}
\usepackage{caption}
\usepackage{tikz}
\usepackage{multicol}
\usetikzlibrary{arrows,positioning,calc}
\usepackage{pstool}
\usepackage{textcomp}
\usepackage[hidelinks]{hyperref}
\hypersetup{colorlinks,bookmarksopen,linkcolor=blue,citecolor=blue,urlcolor=blue}
\usepackage{float}
\usepackage{makecell}
\usepackage[export]{adjustbox}
\usepackage{pdfpages}
\usepackage{booktabs} 

\definecolor{myblue}{RGB}{0,115,189}
\definecolor{mygreen}{rgb}{0.19,0.61,0.21}

\newcommand{\bs}[1]{{\boldsymbol{#1}}}
\newcommand{\tr}[1]{{\mathrm{tr\,{#1}}}}
\newcommand{\rank}[1]{{\mathrm{rank\,{#1}}}}
\newcommand{\order}[1]{{\mathrm{order\,{#1}}}}
\newcommand{\mtrx}[1]{{\boldsymbol{\mathsf{#1}}}}
\newcommand{\tensor}[1]{{\boldsymbol{#1}}}
\newcommand{\tensorfour}[1]{{\mathbb{#1}}}
\newcommand{\column}[1]{{\underline{#1}}}
\newcommand{\tcaption}[1]{\footnotesize{{#1}}}

\def\Aop{\operatornamewithlimits{\mathchoice{\vcenter{\hbox{\Large\sf{A}}}}{\vcenter{\hbox{\Large A}}}{\mathsf{A}}{\mathsf{A}}}}
\DeclareFloatingEnvironment{algorithm}
\DeclareRobustCommand\sampleline[1]{%
	\tikz\draw[#1] (0,0) (0,\the\dimexpr\fontdimen22\textfont2\relax)
	-- (2em,\the\dimexpr\fontdimen22\textfont2\relax);%
}
\graphicspath{{figs/}}

\journal{Adv. Model. Simul. Eng. Sci.}

\begin{document}


\begin{frontmatter}
\title{Reduced integration schemes in micromorphic computational homogenization of elastomeric mechanical metamaterials\tnoteref{titlefoot}}
\tnotetext[mytitlenote]{The post-print version of this article is published in \emph{Adv. Model. Simul. Eng. Sci.}, \href{https://doi.org/10.1186/s40323-020-00152-7}{10.1186/s40323-020-00152-7}}

\author[CTU]{Ond\v{r}ej Roko\v{s}\corref{correspondingauthor}} 
\ead{o.rokos@tue.nl}

\author[CTU]{Jan Zeman}
\ead{jan.zeman@fsv.cvut.cz}

\author[CTU]{Martin Do\v{s}k\'{a}\v{r}}
\ead{martin.doskar@fsv.cvut.cz}

\author[UCSD]{Petr Krysl\fnref{myfootnote}}
\ead{pkrysl@ucsd.edu}

\cortext[correspondingauthor]{Corresponding author.}
\fntext[myfootnote]{Currently Visiting Professor at Czech Technical University in Prague}


\address[CTU]{Department of Mechanics, Faculty of Civil Engineering, Czech Technical University in Prague, Th\'{a}kurova~7, 166~29 Prague~6, Czech Republic.}

\address[UCSD]{Structural Engineering Department, University of California, La Jolla, 24105, San Diego, USA.}


\begin{abstract}
	
Exotic behaviour of mechanical metamaterials often relies on an internal transformation of the underlying microstructure triggered by its local instabilities, rearrangements, and rotations. Depending on the presence and magnitude of such a transformation, effective properties of a metamaterial may change significantly. To capture this phenomenon accurately and efficiently, homogenization schemes are required that reflect microstructural as well as macro-structural instabilities, large deformations, and non-local effects. To this end, a micromorphic computational homogenization scheme has recently been developed, which employs the particular microstructural transformation as a non-local mechanism, magnitude of which is governed by an additional coupled partial differential equation.
Upon discretizing the resulting problem it turns out that the macroscopic stiffness matrix requires integration of macro-element basis functions as well as their derivatives, thus calling for higher-order integration rules. Because evaluation of a constitutive law in multiscale schemes involves an expensive solution of a non-linear boundary value problem, computational efficiency of the micromorphic scheme can be improved by reducing the number of integration points.
Therefore, the goal of this paper is to investigate reduced-order schemes in computational homogenization, with emphasis on the stability of the resulting elements. In particular, arguments for lowering the order of integration from expensive mass-matrix to a cheaper stiffness-matrix equivalent are outlined first. An efficient one-point integration quadrilateral element is then introduced and a proper hourglass stabilization is discussed.
Performance of the resulting set of elements is finally tested on a benchmark bending example, showing that we achieve accuracy comparable to the full quadrature rules, whereas computational cost decreases proportionally to the reduction in the number of quadrature points used.

\end{abstract}


\end{frontmatter}


%
%
\section*{Introduction}
\label{sect:introduction}
\begin{figure}
	\centering
	\begin{tabular}{@{}cc@{}}
		\includegraphics[height=2.2cm]{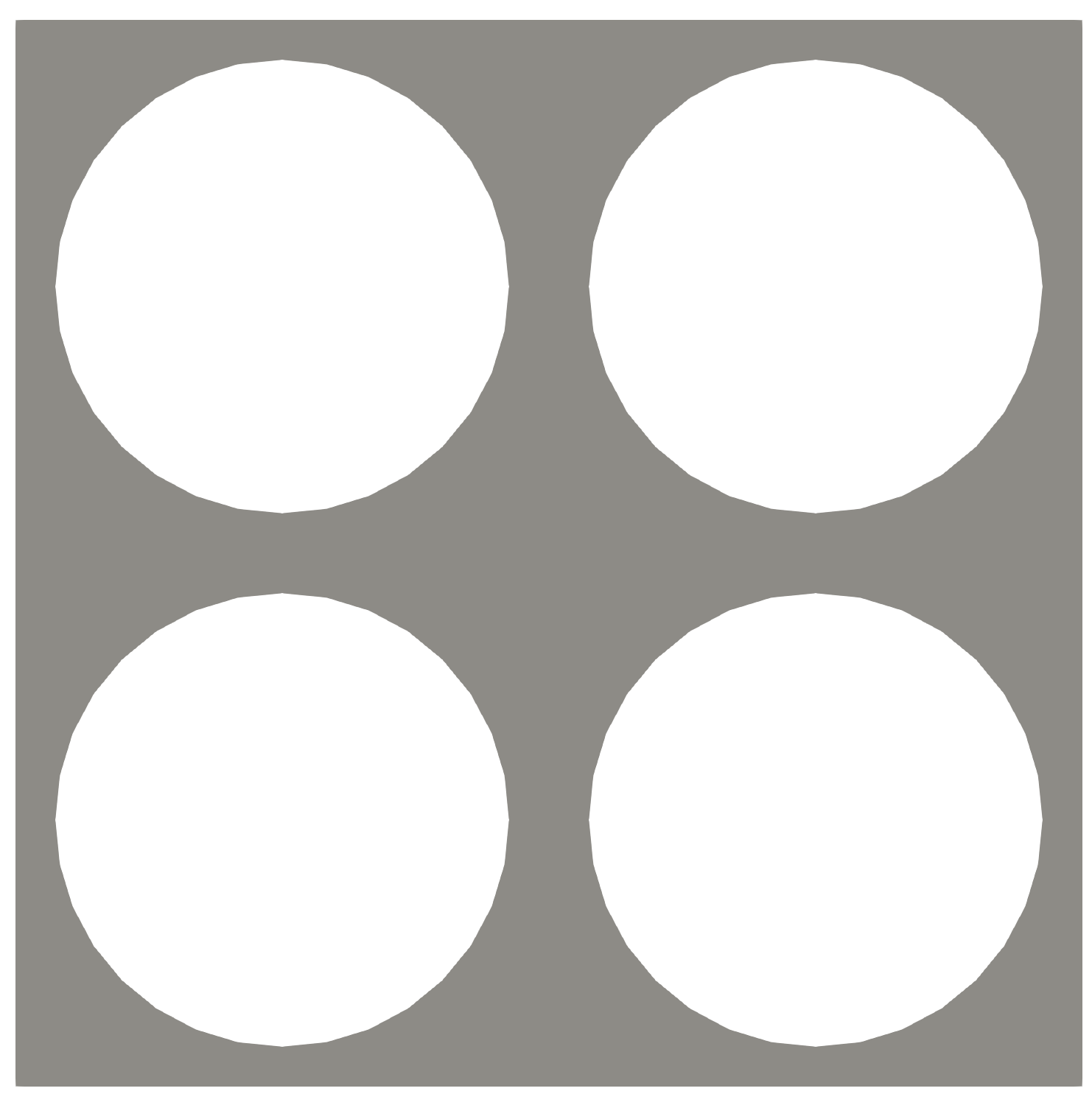}\hspace{1em}\includegraphics[height=2.2cm]{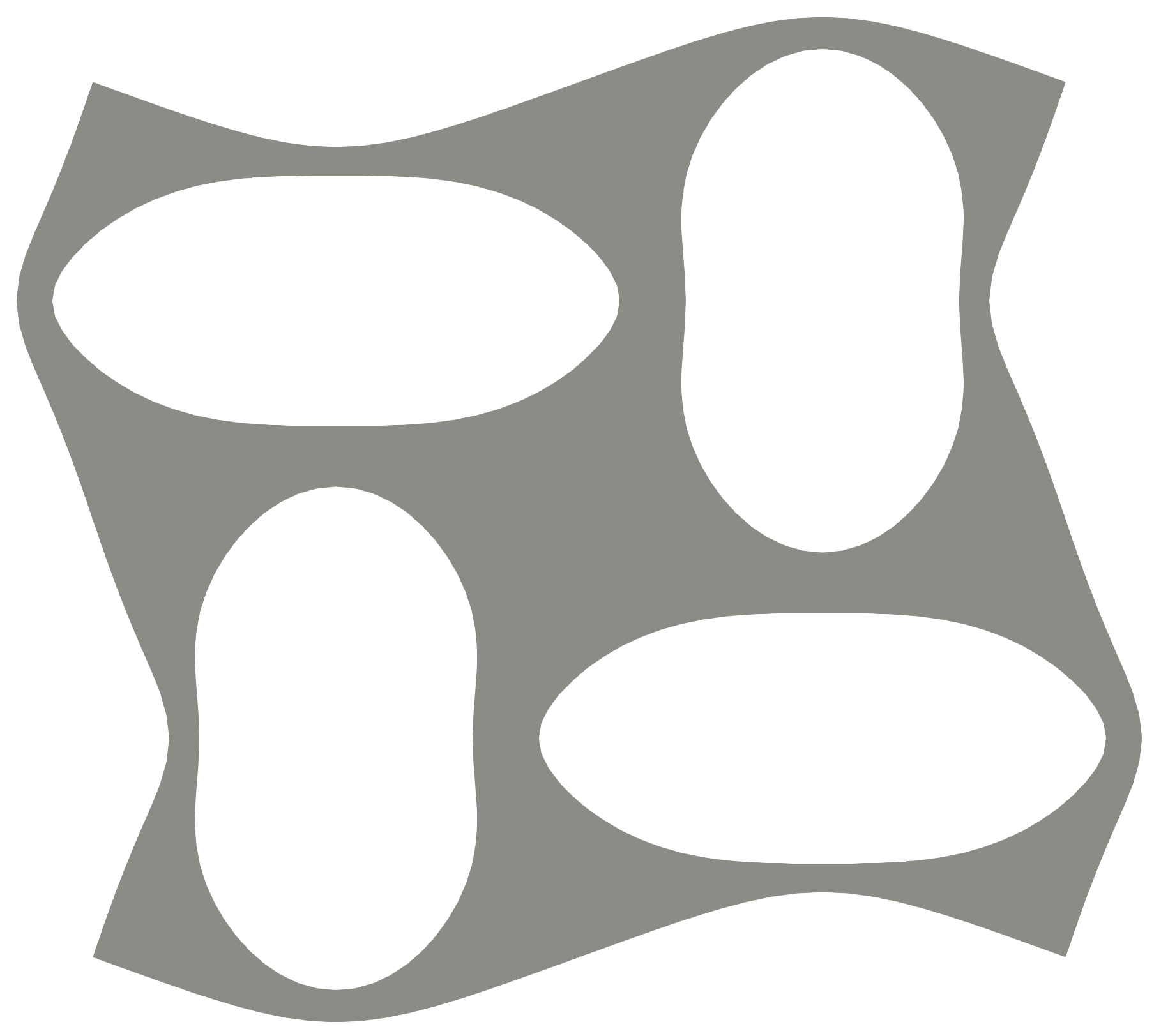} &
		\includegraphics[width=6cm]{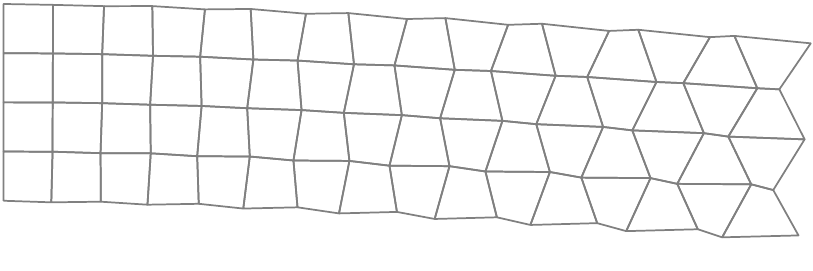} \\
		\tcaption{(a) patterning} & 
		\tcaption{(b) hourglass instability}
	\end{tabular}
	\caption{(a)~Typical patterning mode in mechanical metamaterial consisting of a square stacking of holes, cf.~\cite{Bertoldi2008d}. (b)~Hourglassing observed in under-integrated elements in standard finite element technologies.}
	\label{fig:intro}
\end{figure}

Mechanical metamaterials have recently received a great amount of attention in the engineering literature, aiming at applications ranging from acoustics to soft robotics~\cite{Yang:2015,Mark:2016,Mirzaali:2018,Whitesides:2018}. In the particular case of elastomeric mechanical metamaterials, effective properties often depend on the internal state of the microstructure, dictated by the so-called pattern transformation occurring upon microstructural buckling, see~e.g.~\cite{Bertoldi2008d}, an example of which is shown in Fig.~\ref{fig:intro}a. Such patterning results in abrupt changes in the effective material and structural behaviour, from which non-local effects emerge.

For engineering design and optimization of mechanical metamaterials, efficient numerical modelling tools are required. One of the available options is computational homogenization~\cite{Kouznetsova:2001}, which replaces the effective constitutive behaviour of the macroscopic material with a mechanical system, specified by a Representative Volume Element~(RVE). Advantages of such an approach are that all microstructural features are taken into account, including microstructural morphology, non-linear material behaviour, or local microstructural buckling. However, the macroscopic continuum is treated as local, i.e.~it ignores any communication among neighbouring points (a consequence of the assumption of the formulation on separation of scales). To alleviate this limitation, various enriched theories have been proposed in the literature, including higher-order and micromorphic computational homogenization schemes, see~e.g.~\cite{Kouznetsova2002,Kouznetsova2004,Biswas:2017}. In both first- and higher-order homogenization schemes, the evaluation of the effective constitutive law represents the most computationally intensive operation: it involves a separate non-linear Finite Element~(FE) analysis on the RVE level, from which average quantities such as the homogenized stress and constitutive tangent stiffness are identified. Multiple approaches can be used to reduce the computational effort, including reduced-order modelling at the level of each RVE or considering equivalent surrogate models, see e.g.~\cite{Yvonnet:2013,OLIVER:2017,vanTuijl:2019}.

In this contribution, we focus on a yet another approach to reduce the computational effort in the context of the micromorphic computational homogenization~\cite{Rokos2018} by decreasing the number of macroscopic integration points. We expect that computing times will scale approximately linearly with the number of macroscopic integration points used, since most of the computing effort in a multiscale computational homogenization analysis is spent on evaluating the macro-scale constitutive law from the solutions of individual RVE boundary value problems. Depending on the bandwidth and size of the discretized macroscopic stiffness matrix, a certain overhead due to a repetitive solution of the resulting system of linear equations occurs as well. Furthermore, we systematically test and review performance of several element types. In standard FE technology, in addition to the reduction in computational costs, the under-integration is frequently used to remove artificial stiffening that can be especially severe in the limit of incompressibility and bending-dominated simulations (i.e.~volumetric and shear locking). Within the context of the micromorphic computational homogenization, however, the efficiency is the primal motivation, as employed microstructures are typically heterogeneous and porous, and hence no locking occurs.

Because the stiffness matrix occurring in the discretized version of the micromorphic formulation contains integrals of basis shape functions as well as their derivatives, full integration entails expensive integration rules corresponding to mass matrix equivalents. On the basis of a linear analysis it can be conjectured that integration rules accurate for basis functions' derivatives are sufficient to maintain the proper rank of element stiffness matrices. Furthermore, we demonstrate that a one-integration point quadrilateral, which suffers from the well-known hourglassing, see e.g.~\cite{Belytschko:1984} and Fig.~\ref{fig:intro}b, can be enhanced with standard stabilization procedures and used in micromorphic computational homogenization. 

Throughout the manuscript, scalars are denoted~$a$, vectors~$\vec{a}$, second-order tensors~$\tensor{\sigma}$, fourth-order tensors~$\tensorfour{C}$, column matrices~$\column{a}$, and matrices~$\mtrx{A}$. A position vector in the reference configuration (in two dimensions) is denoted~$ \vec{X} = X_1\vec{e}_1 + X_2\vec{e}_2 $, a gradient operator is defined as~$ \vec{\nabla} \vec{a} = \textstyle \frac{\partial a_j}{\partial X_i} \vec{e}_i \vec{e}_j $, single contraction~$\tensor{A}\cdot\tensor{B} = A_{ik}B_{kj}\vec{e}_i\vec{e}_j $, double contraction~$\tensor{A}:\tensor{B} = A_{ij}B_{ji}$, and right transposition of a fourth-order tensor~$\tensorfour{C} = C_{ijkl}$ as~$\tensorfour{C}^\mathsf{RT} = C_{ijlk}$. Summation over repeated indices is assumed for tensor operations, unless indicated otherwise.
%
%
\section*{Governing equations and discretization}
\label{sect:theory}
%
%
\subsection*{Kinematic decomposition and governing equations}
\label{sect:governing_equations}
The micromorphic computational homogenization, developed originally in~\cite{Rokos2018} and extended later on in~\cite{Rokos2019}, relies on a multiscale kinematic decomposition of the displacement field~$\vec{u}(\vec{X},\vec{X}_\mathrm{m})$ in the close vicinity of a macroscopic point~$\vec{X}$ in the following form
\begin{equation}
\begin{aligned}
\vec{{u}}(\vec{X},\vec{X}_\mathrm{m}) &\approx \vec{v}_0(\vec{X}) + \vec{X}_\mathrm{m}\cdot\vec{\nabla}\vec{v}_0(\vec{X}) \\
&+
\sum_{i=1}^{n} [v_i(\vec{X}) + \vec{X}_\mathrm{m}\cdot\vec{\nabla}v_i(\vec{X})]\vec{\varphi}_i(\vec{X}_\mathrm{m}) \\
&+
\vec{w}(\vec{X},\vec{X}_\mathrm{m}).
\end{aligned}
\label{eq:approxu}
\end{equation}
Two scales are present, a macroscopic scale, spanned by~$\vec{X} \in \Omega$, and a local microscopic scale, spanned by~$\vec{X}_\mathrm{m} \in \Omega_\mathrm{m}$, where~$\Omega$ denotes the macroscopic domain and~$\Omega_\mathrm{m}$ a microscopic RVE associated with each macroscopic point~$\vec{X}$. The unknown fields to be computed are the macroscopic mean solution~$\vec{v}_0$, micromorphic fields~$v_i$, $i = 1,\dots,n$, that act as magnitudes of patterning modes~$\vec{\varphi}_i$, and a microfluctuation field~$\vec{w}$. Individual patterning modes~$\vec{\varphi}_i$, $i = 1, \dots, n$, are assumed to be known a priori, identified either experimentally~\cite{Siavash:2020} or through a buckling or Floquet--Bloch analysis~\cite{Bertoldi2008d}. An example of a patterning mode in an infinite microstructure with a square stacking of holes is shown in Fig.~\ref{fig:intro}a.

Assuming a hyperelastic behaviour of the elastomeric base material, specified by an elastic energy density function~$W$, an averaged total potential energy~$\mathcal{E}$ can be defined as
\begin{equation}
\mathcal{E}(\vec{u}) = \frac{1}{|\Omega_\mathrm{m}|} \int_\Omega \int_{\Omega_\mathrm{m}} W(\vec{X},\tensor{F}) \, \mathrm{d}\vec{X}_\mathrm{m}\mathrm{d}\vec{X}.
\label{eq:energy}
\end{equation}
Dropping for simplicity of the explanation any additional constraints imposed on the~$\vec{w}$ term (specified later in Eqs.~\eqref{eq:mgoverningc}--\eqref{eq:mgoverningf}), the first variation of~$\mathcal{E}$ reads
\begin{equation}
\delta\mathcal{E}(\vec{u};\delta\vec{u}) = \frac{1}{|\Omega_\mathrm{m}|} \int_\Omega \int_{\Omega_\mathrm{m}} \tensor{P} : \vec{\nabla}_\mathrm{m}\delta\vec{u}(\vec{X},\vec{X}_\mathrm{m}) \, \mathrm{d}\vec{X}_\mathrm{m}\mathrm{d}\vec{X},
\label{eq:firstVariation}
\end{equation}
where~$\tensor{P}(\tensor{F}) = \frac{\partial W(\tensor{F})}{\partial{\tensor{F}}^\mathsf{T}}$ is the microscopic first Piola--Kirchhoff stress tensor, $\vec{\nabla}_\mathrm{m} = \frac{\partial}{\partial X_{\mathrm{m},i}}$ denotes microscopic gradient operator with respect to the reference configuration, $\vec{u}$ is a function of all unknown field quantities (i.e.~$\vec{v}_0$, $v_i$, and~$\vec{w}$), and~$|\Omega_\mathrm{m}|$ stands for the RVE area. Making use of the kinematic decomposition of Eq.~\eqref{eq:approxu}, a set of macroscopic and microscopic governing equations can be derived following the standard rules of the calculus of variations, see~\cite{Rokos2018,Rokos2019} for further details.

The mean field~$\vec{v}_0$ is found to be governed by the following balance equation of linear momentum
\begin{align}
\vec{\nabla}\cdot\bs{\Theta}^\mathsf{T} &= \vec{0}, \quad \mbox{in}\ \Omega,\label{eq:v0a} \\
\bs{\Theta}\cdot\vec{N} &= \vec{0}, \quad \mbox{on}\ \Gamma_\mathrm{N}, \label{eq:v0b}
\end{align}
where~$\vec{\nabla} = \frac{\partial}{\partial X_i}$ denotes the macroscopic gradient operator, $\vec{N}$ is the outer unit normal to the boundary of the macroscopic domain~$\partial\Omega$, and~$\Gamma_\mathrm{N} \subseteq \partial\Omega$ is the part of the macroscopic domain where zero tractions are prescribed. The magnitude~$v_i$ of each patterning field is governed by the following scalar partial differential equation
\begin{align}
\vec{\nabla}\cdot \vec{\Lambda}_{i}-\Pi_{i} &= 0, \quad \mbox{in}\ \Omega, &&i = 1, \dots, n \label{eq:via}\\
\vec{\Lambda}_{i}\cdot\vec{N} &= 0, \quad \mbox{on}\ \Gamma_\mathrm{N}, &&i = 1, \dots, n. \label{eq:vib}
\end{align}
In Eqs.~\eqref{eq:v0a}--\eqref{eq:vib}, the homogenized stress quantities read as
\begin{align}
\bs{\Theta} &= \frac{1}{|\Omega_\mathrm{m}|}\int_{\Omega_\mathrm{m}} \bs{P} \, \mathrm{d}\vec{X}_\mathrm{m}, \label{fe2:eq10a}\\
\Pi_{i} &= \frac{1}{|\Omega_\mathrm{m}|}\int_{\Omega_\mathrm{m}} \bs{P}:\vec{\nabla}_\mathrm{m}\vec{\varphi}_i \, \mathrm{d}\vec{X}_\mathrm{m}, &&i = 1, \dots, n, \label{fe2:eq10b}\\
\vec{\Lambda}_{i} &= \frac{1}{|\Omega_\mathrm{m}|}\int_{\Omega_\mathrm{m}} \bs{P}^\mathsf{T}\cdot\vec{\varphi}_i + \vec{X}_\mathrm{m}[\bs{P}:\vec{\nabla}_\mathrm{m}\vec{\varphi}_i] \, \mathrm{d}\vec{X}_\mathrm{m}, &&i = 1, \dots, n.
\label{fe2:eq10c}
\end{align}

The microfluctuation fields~$\vec{w}(\vec{X},\vec{X}_\mathrm{m})$, $\vec{X}_\mathrm{m} \in \Omega_\mathrm{m}$, associated with each macroscopic point~$\vec{X}$, satisfy the following set of equations
\begin{align}
\vec{\nabla}_\mathrm{m}\cdot\bs{P}^\mathsf{T} &= \vec{\mu} + \sum_{i=1}^{n}\nu_i\vec{\varphi}_i + \sum_{i = 1}^{n}\vec{\eta}_i \cdot (\vec{\varphi}_i\vec{X}_\mathrm{m}), \ &&\mbox{in}\ \Omega_\mathrm{m}, \label{eq:mgoverninga}\\
\bs{P}\cdot\vec{N}_\mathrm{m} &= \pm\vec{\lambda}, \ &&\mbox{on}\ \partial\Omega_\mathrm{m}^\pm, \label{eq:mgoverningb}\\
\vec{0} &= \llbracket \vec{w}(\vec{X},\vec{X}_\mathrm{m}) \rrbracket, \ &&\vec{X}_\mathrm{m} \in \partial\Omega_\mathrm{m}^+,\label{eq:mgoverningc}\\
\vec{0} &= \int_{\Omega_\mathrm{m}} \vec{w}(\vec{X},\vec{X}_\mathrm{m}) \,\mathrm{d}\vec{X}_\mathrm{m}, \label{eq:mgoverningd}\\
0 &= \int_{\Omega_\mathrm{m}} \vec{w}(\vec{X},\vec{X}_\mathrm{m}) \cdot \vec{\varphi}_i(\vec{X}_\mathrm{m})\,\mathrm{d}\vec{X}_\mathrm{m}, \ &&i=1,\dots,n,\label{eq:mgoverninge}\\
\vec{0} &= \int_{\Omega_\mathrm{m}} \vec{w}(\vec{X},\vec{X}_\mathrm{m}) \cdot [ \vec{\varphi}_i(\vec{X}_\mathrm{m}) \vec{X}_\mathrm{m} ]\,\mathrm{d}\vec{X}_\mathrm{m}, \ &&i=1,\dots,n.\label{eq:mgoverningf}
\end{align}
In the governing equation~\eqref{eq:mgoverninga}, $\mu$, $\nu_i$, and~$\vec{\eta}_i$ denote the Lagrange multipliers associated with the orthogonality constraints imposed on~$\vec{w}$, Eqs.~\eqref{eq:mgoverningd}--\eqref{eq:mgoverningf}, $\vec{\lambda}$ is the Lagrange multiplier (equivalent to RVE boundary tractions) enforcing the periodicity constraint of Eq.~\eqref{eq:mgoverningb}, and $\llbracket \vec{w}(\vec{X},\vec{X}_\mathrm{m}) \rrbracket = \vec{w}(\vec{X},\partial\Omega_\mathrm{m}^+) - \vec{w}(\vec{X},\partial\Omega_\mathrm{m}^-)$ denotes the jump of the field~$\vec{w}(\vec{X},\vec{X}_\mathrm{m})$ across the RVE boundary~$\partial\Omega_\mathrm{m} = \partial\Omega_\mathrm{m}^+ \cup \partial\Omega_\mathrm{m}^-$ split into two parts, image~$\Omega_\mathrm{m}^+$ and mirror~$\Omega_\mathrm{m}^-$ (distinguished in Fig.~\ref{fig:geometry}b in green and blue colour).

Full details on the derivation of macro- and micro-scopic governing equations and energy considerations are available in~\cite{Rokos2018,Rokos2019}, whereas comparison of the micro-morphic scheme with first- and second-order computational homogenization is provided in~\cite{Sven:2019}.
%
%
\subsection*{Solution procedure and discretization}
\label{sect:discretizaion}
Using the standard FE procedures as described in~\cite{Bree2019}, the macroscopic governing equations can be discretized, resulting in the iterative system of macroscopic equations
\begin{equation}
\mtrx{K}_\mathrm{M}\Delta\column{v} = \column{f}_\mathrm{ext} - \column{f}_\mathrm{M},
\label{eq:Mnewton}
\end{equation}
where~$\Delta\column{v} = [\Delta\column{v}_0,\dots,\Delta\column{v}_n]^\mathsf{T}$ denotes the increment of macroscopic Degrees Of Freedom~(DOFs), $\column{f}_\mathrm{ext}$ represents an external loading applied to~$\column{v}_0$, $\column{f}_\mathrm{M} = \Aop_{e=1}^{n_e} \column{f}_\mathrm{M}^e $ is the vector of internal macroscopic forces, $\mtrx{K}_\mathrm{M} = \Aop_{e=1}^{n_e} \mtrx{K}_\mathrm{M}^e$ is the global macroscopic stiffness matrix, and~$\Aop$ denotes the assembly operator. 

Internal forces for each macroscopic element~$e = 1,\dots,n_e$ are computed as
\begin{align}
\column{f}_\mathrm{M}^e 
&= [\column{f}_0^e, \dots, \column{f}_n^e]^\mathsf{T}, \label{eq:Mfint} \\
\column{f}_0^e &= \sum_{i_\mathrm{g}=1}^{n_\mathrm{g}}w_{i_\mathrm{g}}J_{i_\mathrm{g}}\big(\mtrx{B}_0^\mathsf{T}\column{\Theta}\big), \\
\column{f}_i^e &= \sum_{i_\mathrm{g}=1}^{n_\mathrm{g}}w_{i_\mathrm{g}}J_{i_\mathrm{g}}\big(\mtrx{B}_i^\mathsf{T}\column{\Lambda}_i - \mtrx{N}_i^\mathsf{T}\Pi_i\big), \quad i = 1,\dots,n,
\end{align}
where~$\mtrx{N}_\bullet$ are the standard matrices of the macroscopic shape interpolation functions, $\mtrx{B}_\bullet$ stand for their derivatives, and~$w_{i_\mathrm{g}}$ are integration weights with the corresponding Jacobians~$J_{i_\mathrm{g}}$. The column matrices~$\column{\Theta}$ and~$\column{\Lambda}_i$ store components of the homogenized stress quantities defined in Eqs.~\eqref{fe2:eq10a} and~\eqref{fe2:eq10c}. 

The macroscopic single-element stiffness matrix can be expressed as
\begin{equation}
\begin{aligned}
\mtrx{K}^e_\mathrm{M}
&= \begin{bmatrix}
\mtrx{K}_{00}^e & \dots & \mtrx{K}_{0n}^e \\
\vdots & \ddots & \vdots\\
\mtrx{K}_{n0}^e & \dots &\mtrx{K}_{nn}^e
\end{bmatrix} \\
& =
\sum_{i_\mathrm{g} = 1}^{n_\mathrm{g}}w_{i_\mathrm{g}}J_{i_\mathrm{g}}
\left(
\underbrace{
	\begin{bmatrix}
	\mtrx{D}_{00} & \dots & \mtrx{D}_{0n} \\
	\vdots & \ddots & \vdots\\
	\mtrx{D}_{n0} & \dots &\mtrx{D}_{nn}
	\end{bmatrix}}_{\mtrx{H}}
-
\begin{bmatrix}
\mtrx{D}_{0w} \\
\vdots\\
\mtrx{D}_{nw}
\end{bmatrix}
\begin{bmatrix}
\mtrx{D}_{ww}
\end{bmatrix}^{-1}
\begin{bmatrix}
\mtrx{D}_{w0} \dots \mtrx{D}_{wn}
\end{bmatrix}\right),
\end{aligned}
\label{eq:Mk}
\end{equation}
where the first term on the right-hand side reflects stiffness quantities obtained by volume averaging of microscopic constitutive tangent~$\tensorfour{C}(\tensor{F}) = \frac{\partial \tensor{P}(\tensor{F})}{\partial{\tensor{F}}^\mathsf{T}}$ weighted by~$\vec{\varphi}_i$, $\vec{\nabla}_\mathrm{m}\vec{\varphi}_i$, or~$\vec{X}_\mathrm{m}\vec{\nabla}_\mathrm{m}\vec{\varphi}_i$, which can be derived from the second variation of the averaged total potential energy~$\mathcal{E}$, i.e.
\begin{equation}
\delta^2\mathcal{E}(\vec{u};\delta\vec{u}) = \frac{1}{|\Omega_\mathrm{m}|} \int_\Omega \int_{\Omega_\mathrm{m}} \vec{\nabla}_\mathrm{m}\delta\vec{u}(\vec{X},\vec{X}_\mathrm{m}) : \tensorfour{C} : \vec{\nabla}_\mathrm{m}\delta\vec{u}(\vec{X},\vec{X}_\mathrm{m}) \, \mathrm{d}\vec{X}_\mathrm{m}\mathrm{d}\vec{X},
\label{eq:secondVariation}
\end{equation}
while making use of the kinematic decomposition of Eq.~\eqref{eq:approxu}. The second term of Eq.~\eqref{eq:Mk} corresponds to the Schur complement of the microfluctuation field~$\vec{w}$ solved for at the microscale, and thus coupled to all macroscopic quantities. Note that~$\mtrx{D}_{ww} = \frac{1}{|\Omega_\mathrm{m}|}\mtrx{H}_{ww}$, where~$\mtrx{H}_{ww}$ is defined in Eq.~\eqref{eq:mnewton} below, and~$\mtrx{D}_{iw}$, $i = 0, \dots, n$, are coupling terms between the macroscopic and microfluctuation fields, not discussed here in detail for brevity. The stiffness density term associated with the mean field~$\vec{v}_0$ has the standard form, that is
\begin{equation}
\mtrx{D}_{00} = \mtrx{B}_0^\mathsf{T}\overline{\mtrx{C}}\mtrx{B}_0, \quad\mbox{with}\quad \overline{\mtrx{C}} = \frac{1}{|\Omega_\mathrm{m}|} \int_{\Omega_\mathrm{m}} \mtrx{C} \,\mathrm{d}\vec{X}_\mathrm{m},
\label{eq:K00}
\end{equation}
where~$\mtrx{C}$ stands for the matrix representation of the constitutive tangent~$\tensorfour{C}$. The micromorphic stiffness densities have the following structure
\begin{align}
\mtrx{D}_{0i} &= \mtrx{D}_{i0}^\mathsf{T} = \mtrx{B}_0^\mathsf{T}\overline{\mtrx{E}}_{0i}\mtrx{B}_i + \mtrx{B}_0^\mathsf{T}\overline{\column{F}}_{0i}\mtrx{N}_i, &i &= 1,\dots,n, \label{eq:Kiia} \\
\mtrx{D}_{ij} &= \underbrace{\mtrx{B}_i^\mathsf{T}\overline{\mtrx{E}}_{ij}\mtrx{B}_j}_{\mtrx{U}_{ij}} +
\mtrx{B}_i^\mathsf{T}\overline{\column{F}}_{ij}\mtrx{N}_j +
\mtrx{N}_i^\mathsf{T}\overline{\column{F}}_{ij}^\mathsf{T}\mtrx{B}_j +
\underbrace{\mtrx{N}_i^\mathsf{T}\overline{G}_{ij}\mtrx{N}_j}_{\mtrx{V}_{ij}}, &i,\,j &= 1,\dots,n,
\label{eq:Kii}
\end{align}
where no summation on repeated indices~$i$ and~$j$ is implied. The averaged constitutive tangents~$\overline{\mtrx{E}}_{ij}$, $\overline{\column{F}}_{ij}$, and~$\overline{G}_{ij}$ are obtained as more involved weighted averages of the microscopic constitutive tensor~$\tensorfour{C}$ over the RVE domain, recall Eq.~\eqref{eq:secondVariation}, expressed explicitly in tensor notation as
\begingroup
\allowdisplaybreaks
\begin{align}
\overline{\tensor{E}}_{0i} &= \frac{1}{|\Omega_\mathrm{m}|} \int_{\Omega_\mathrm{m}} \tensorfour{C}^\mathsf{RT}\cdot\vec{\varphi}_i + [(\tensorfour{C}:\vec{\nabla}_\mathrm{m}\vec{\varphi}_i) \vec{X}_\mathrm{m}] \,\mathrm{d}\vec{X}_\mathrm{m},
\label{eq:Mstiffnessesa}\\
\overline{\tensor{F}}_{0i} &= \frac{1}{|\Omega_\mathrm{m}|} \int_{\Omega_\mathrm{m}} \tensorfour{C}:\vec{\nabla}_\mathrm{m}\vec{\varphi}_i \,\mathrm{d}\vec{X}_\mathrm{m}, 
\label{eq:Mstiffnessesb}\\
\overline{\tensor{E}}_{ij} &= \frac{1}{|\Omega_\mathrm{m}|} \int_{\Omega_\mathrm{m}}
\vec{\varphi}_i \cdot \tensorfour{C}^\mathsf{RT} \cdot \vec{\varphi}_j +
(\vec{\varphi}_i \cdot \tensorfour{C} : \vec{\nabla}_\mathrm{m}\vec{\varphi}_j) \vec{X}_\mathrm{m} \nonumber\\
&+ \vec{X}_\mathrm{m}(\vec{\nabla}_\mathrm{m}\vec{\varphi}_i : \tensorfour{C}^\mathsf{RT} \cdot \vec{\varphi}_j) + \vec{X}_\mathrm{m}(\vec{\nabla}_\mathrm{m}\vec{\varphi}_i : \tensorfour{C} : \vec{\nabla}_\mathrm{m}\vec{\varphi}_j) \vec{X}_\mathrm{m} \,\mathrm{d}\vec{X}_\mathrm{m},\label{eq:Mstiffnessesc} \\
\vec{\overline{F}}_{ij} &= \frac{1}{|\Omega_\mathrm{m}|} \int_{\Omega_\mathrm{m}}
\vec{\nabla}_\mathrm{m}\vec{\varphi}_i 
: [ \, \tensorfour{C}^\mathsf{RT} \cdot\vec{\varphi}_j  
+ (\tensorfour{C} :\vec{\nabla}_\mathrm{m}\vec{\varphi}_j) \vec{X}_\mathrm{m} \, ]\,\mathrm{d}\vec{X}_\mathrm{m},
\label{eq:Mstiffnessesd}\\
\overline{G}_{ij} &= \frac{1}{|\Omega_\mathrm{m}|} \int_{\Omega_\mathrm{m}} \vec{\nabla}_\mathrm{m}\vec{\varphi}_i
:
\tensorfour{C}:
\vec{\nabla}_\mathrm{m}\vec{\varphi}_j \,\mathrm{d}\vec{X}_\mathrm{m}, 
\label{eq:Mstiffnessese}
\end{align}
\endgroup
see~\cite{Bree2019} for more details. Note that the matrices of basis interpolation functions~$\mtrx{N}_i$ as well as their derivatives~$\mtrx{B}_i$, $i = 0,\dots,n$, in Eqs.~\eqref{eq:K00}--\eqref{eq:Kii} depend only on the macroscopic spatial variable~$\vec{X}$, whereas all quantities related to material properties (that is~$\column{\Theta}$, $\column{\Lambda}_{i}$, $\Pi_i$, $\overline{\mtrx{C}}$, $\overline{\mtrx{E}}_{ij}$, $\overline{\column{F}}_{ij}$, and~$\overline{G}_{ij}$) depend in a non-linear manner on the micro-fluctuations~$\vec{w}(\vec{X},\vec{X}_\mathrm{m})$.

Finally, at each macroscopic quadrature point, $i_\mathrm{g}$, a microscopic problem for~$\column{w}$ is iteratively solved from
\begin{equation}
\underbrace{
	\begin{bmatrix}
	\mtrx{K}_{ww} & \mtrx{A}^\mathsf{T} \\
	\mtrx{A} & \mtrx{0}
	\end{bmatrix}
}_{\mtrx{H}_{ww}}
\begin{bmatrix}
\Delta\underline{w}\\
\underline{\lambda}
\end{bmatrix}
=
\begin{bmatrix}
\underline{f}_w\\
\column{0}
\end{bmatrix},
\label{eq:mnewton}
\end{equation}
where~$\mtrx{K}_{ww}$ is the microscopic stiffness matrix, $\column{f}_w$ stores microscopic internal forces, $\column{\lambda}$ collects all discretized Lagrange multipliers, and~$\mtrx{A}$ is a constraint matrix reflecting all equality constraints required for~$\vec{w}$, listed in Eqs.~\eqref{eq:mgoverningc}--\eqref{eq:mgoverningf}.

The overall solution procedure is summarized as follows. For each macroscopic Gauss integration point~$i_\mathrm{g}$, an associated non-linear microscopic RVE problem of Eq.~\eqref{eq:mnewton} is solved for~$\vec{w}$. When a converged~$\vec{w}$ is achieved, the microscopic first Piola--Kirchhoff stress tensor~$\tensor{P}$ and constitutive tangent~$\tensorfour{C}$ can be established in the RVE domain, and the homogenized stresses~\eqref{fe2:eq10a}--\eqref{fe2:eq10c} along with corresponding stiffnesses~\eqref{eq:K00}--\eqref{eq:Mstiffnessese} follow by their averaging. The macroscopic internal forces~\eqref{eq:Mfint} and stiffness matrices~\eqref{eq:Mk} are subsequently integrated, and the resulting macroscopic system~\eqref{eq:Mnewton} assembled and solved. Since all microfluctuation fields~$\column{w}$ are condensed out (i.e.~solved at the Gauss integration point level), all macroscopic quantities~$\column{v}_0$ and~$\column{v}_i$ directly follow from Eq.~\eqref{eq:Mnewton}. Further details on discretization and numerical implementation of the micromorphic computational homogenization scheme can be found in~\cite{Bree2019,Sylvia:thesis}.
%
%
\section*{Numerical integration and stabilization}
\label{sect:stabilization}
%
%
\subsection*{Full integration}
\label{sect:full_integration}
It may appear from the structure of~$\mtrx{D}_{ij}$ in Eq.~\eqref{eq:Kii} that for the full integration of the macroscopic element stiffness matrix associated with the micromorphic field a mass-matrix equivalent integration rule should be employed, i.e.~a rule exactly integrating the~$\int_\Omega \mtrx{N}^\mathsf{T} \mtrx{N} \, \mathrm{d}\vec{X}$ term for an undistorted nor mapped element, recalled for convenience in Tab.~\ref{tab:integration} for a few typical element types. Note that the listed quadrature rules are accurate for polynomial functions. The integrated quantities in FEM are, however, rational functions in general. Such integration rules thus might represent a sub-optimal choice, yet they are often employed in numerical simulations for non-linear problems and isoparametric elements, see~e.g.~\cite[Section~5.5.5]{Bathe:2006}.

Because evaluation of the macroscopic constitutive law is a rather expensive operation involving a separate solution of a non-linear FEM system~\eqref{eq:mnewton}, it is desirable to reduce the number of macroscopic integration points. This may be achieved in two ways: 
\begin{enumerate}\setlength{\itemsep}{0pt}\setlength{\parskip}{0pt}\setlength{\parsep}{0pt}
	\item[(i)] using a lower-order interpolation scheme for the micromorphic fields~$v_i$ than for the mean field~$\vec{v}_0$ (and thus requiring a lower order of integration);
	
	\item[(ii)] under-integrating the element stiffness expression while keeping the same interpolation scheme for~$\vec{v}_0$ and~$v_i$.
\end{enumerate}
To achieve high approximation quality of all involved fields, especially in the regime of small separation of scales, we opt for the latter alternative, although from the standard asymptotic homogenization theory one might expect that the magnitude of the fluctuation~$v_i$ would scale with the gradient of~$\vec{v}_0$ in the limit of infinite scale separation, and hence lower approximation space might suffice. Note that because a standard Cauchy continuum is used at the microscale, a standard FE discretization with a full integration scheme is used to discretize the microscopic balance equations~\eqref{eq:mgoverninga}--\eqref{eq:mgoverningf}.
\begin{table}
	\begin{minipage}[t]{\linewidth}
		\let\footnoterule\relax
		\centering
		\caption{The number of Gauss integration points for considered element types according to~\cite[Fig.~6.22]{ZIENKIEWICZ:2013}, \cite[Tab.~5.9]{Bathe:2006}. T3 denotes three-node linear triangle, T6 six-node quadratic triangle, Q4 four-node bilinear quadrilateral, and Q8 eight-node serendipity quadratic quadrilateral. Numbers in parentheses specify the number of Gauss integration points for under-integrated versions in standard FE technology.}
		\renewcommand*{\arraystretch}{1.3}
		\label{tab:integration}
		\begin{tabular}{lcccc}\hline
			Element type & T3 & T6 & Q4 & Q8 \\\hline
			$\mtrx{B}^\mathsf{T}\overline{\mtrx{E}}\mtrx{B}$ (stiffness-matrix) & 1 & 3 & 4 (1\footnote{In standard two-dimensional FEM, two spurious singular modes requiring hourglass control are observed.}) & 9 (4\footnote{Induces one spurious singular mode in standard FEM; the mode is non-communicable, i.e.~it typically does not occur in an element patch and hence no stabilization is usually required, cf. Fig~\ref{fig:modes}a, \cite[Section~5.5.7.]{Bathe:2006}, and~\cite{Fan:1992}.}) \\
			$\mtrx{N}^\mathsf{T}\overline{\mtrx{G}}\mtrx{N}$ (mass-matrix) & 3 & 6 & 4 & 9 \\ \hline
		\end{tabular}
	\end{minipage}
\end{table}
%
%
\subsection*{Reduced integration and spurious singular modes}
\label{sect:reduced_integration}
To conjecture that the order of integration can be lowered from the mass-matrix type of quadrature to the stiffness-matrix equivalent without introducing any additional spurious singular modes, let us start with a set of supporting arguments carried out for a single element stiffness matrix considered in the reference configuration (i.e.~in the limit of small deformations).

For the sake of argument we assume the penalty method being used instead of the equivalent Lagrange multiplier formulation, merely to avoid indefiniteness of~$\mtrx{D}_{ww}$ in Eq.~\eqref{eq:Mk}. Assuming further that the employed mechanical system is stable in the reference configuration, i.e. the corresponding stiffness matrix is positive definite (upon suppressing proper physical singular modes, i.e.~Rigid Body Modes, abbreviated as RBM in what follows), one can argue by means of~\cite[Eq.~(7.7.5)]{Horn:2012} that the Schur complement in Eq.~\eqref{eq:Mk} will not introduce additional spurious singular modes, hence it can be neglected in further considerations. From central symmetry of the employed RVE, $\Omega_\mathrm{m}$, and the associated patterning modes, $\vec{\varphi}_i$, it follows (or can be numerically verified) that quantities~$\overline{\mtrx{E}}_{ij}$, $\overline{\column{F}}_{ij}$, and~$\overline{G}_{ij}$, $i,\,j = 0,\dots,n$, $i\neq j$, in Eqs.~\eqref{eq:Mk}--\eqref{eq:Kii} are zero for square as well as for hexagonal stacking of holes~\cite{Rokos2019}. The matrix~$\mtrx{H}$ is thus block-diagonal and its rank, and therefore also the rank of~$\mtrx{K}_\mathrm{M}^e$, reads
\begin{equation}
\rank(\mtrx{K}_\mathrm{M}^e) \geq \order(\mtrx{D}_{00}) - n_\mathrm{rbm} + \sum_{i=1}^n\rank(\mtrx{D}_{ii}),
\label{eq:rankH}	
\end{equation}
where~$n_\mathrm{rbm}$ is the number of RBM associated with~$\vec{v}_0$, $n$ is the number of micromorphic fields, recall Eq.~\eqref{eq:approxu}, and~$\order(\mtrx{A}) = m$ for a symmetric~$m \times m$ matrix~$\mtrx{A}$. To eliminate any spurious singular modes in the system, $\mtrx{D}_{ii}$ is required to have a full rank, i.e.~$\order(\mtrx{D}_{ii}) = \rank(\mtrx{D}_{ii})$. This condition will be satisfied for both integration rules listed in Tab.~\ref{tab:integration} (i.e.~for stiffness- as well as mass-equivalent, but not for their under-integrated versions in brackets). This is because the constant vector~$\column{1}$ is in all cases considered an element of the kernel of~$\mtrx{U}_{ii}$, whereas it is not in the kernel of~$\mtrx{V}_{ii}$ (recall Eq.~\eqref{eq:Kii} for the definition of~$\mtrx{U}_{ii}$ and~$\mtrx{V}_{ii}$). For the stiffness-equivalent integration rule we thus conclude that the matrix~$\mtrx{D}_{ii}$ maintains its full rank.

\begin{figure}
	\centering
	\begin{tabular}{@{}cc@{}}
		\includegraphics[scale=0.9]{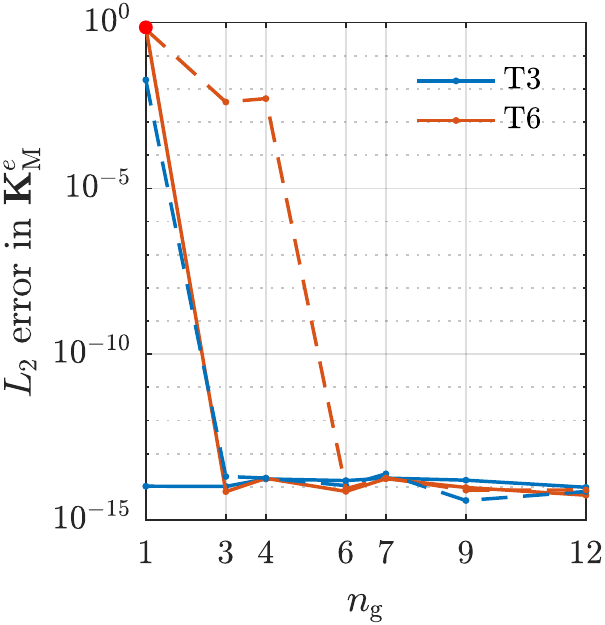} & 
		\includegraphics[scale=0.9]{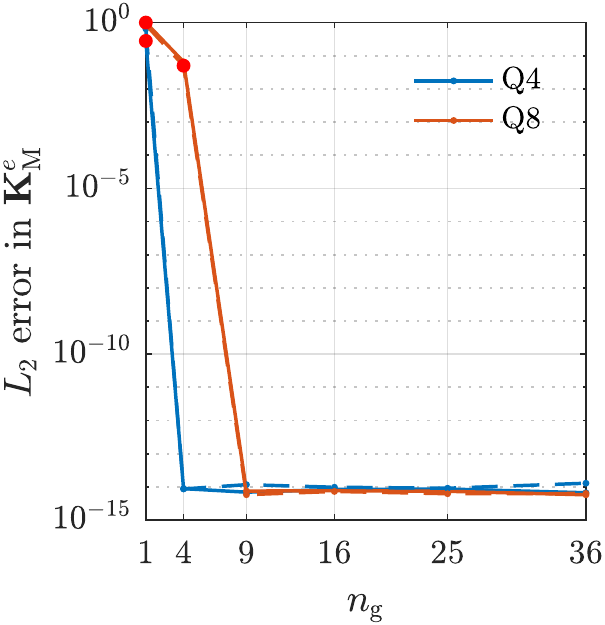} \\
		\tcaption{(a) element stiffness, triangles} &
		\tcaption{(b) element stiffness, quadrangles}
	\end{tabular}
	\caption{$L_2$ errors of a single element stiffness matrix for triangular~(a) and quadrilateral~(b) elements as a function of the number of Gauss integration points. The accuracy is measured relative to the stiffness matrix obtained for the maximum number of integration points used ($n_\mathrm{g} = 13$ for triangles and~$n_\mathrm{g} = 49$ for quadrilaterals). Solid lines (\sampleline{solid}) correspond to~$\mtrx{K}_{00}^e$ and dashed lines (\sampleline{dash pattern=on .7em off .3em}) to~$\mtrx{K}_{11}^e$ sub-matrices of~$\mtrx{K}^e_\mathrm{M}$ in Eq.~\eqref{eq:Mk}, whereas~$\mtrx{K}_{01}^e = (\mtrx{K}_{10}^e)^\mathsf{T} = \mtrx{0}$; elements from Tab.~\ref{tab:integration} are used. Configuration of elements correspond to the reference undeformed state (microstructure with a square stacking of holes used, i.e., $n=1$). Integration rules for which spurious singular modes occur are denoted with red circles.}
	\label{fig:integrationref}
\end{figure}

\begin{figure}
	\centering
	\begin{tabular}{@{}cc@{}}
		\includegraphics[scale=0.9]{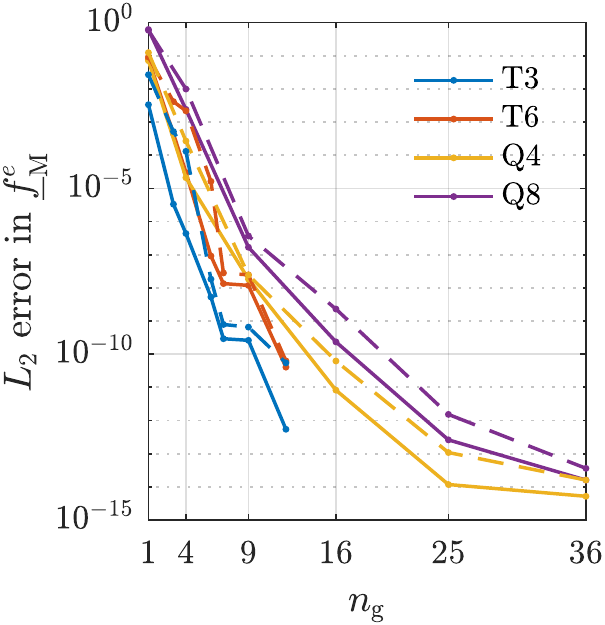} & 
		\includegraphics[scale=0.9]{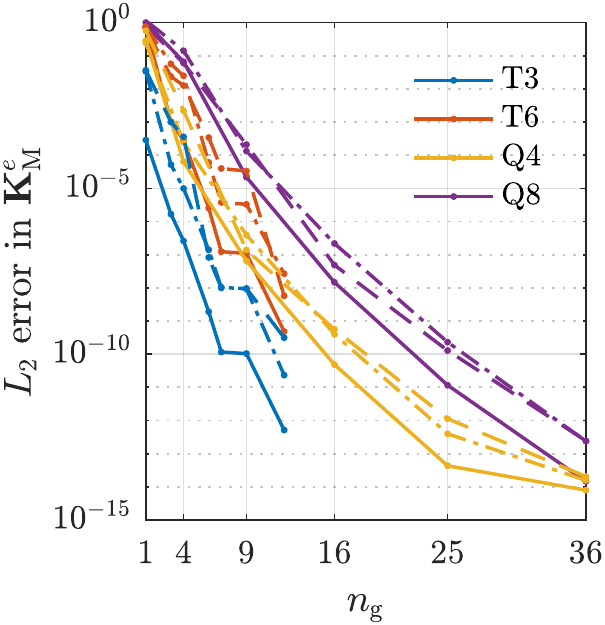} \\
		\tcaption{(a) element internal forces} &
		\tcaption{(b) element stiffness}
	\end{tabular}	
	\caption{$L_2$ errors of a single element internal forces~(a) and stiffnesses~(b) as a function of the number of Gauss integration points. The accuracy is measured relative to the force or stiffness matrix obtained for the maximum number of integration points used ($n_\mathrm{g} = 13$ for triangles and~$n_\mathrm{g} = 49$ for quadrilaterals). Solid lines (\sampleline{solid}) correspond to~$\column{f}_0^e$ or~$\mtrx{K}_{00}^e$, dashed lines (\sampleline{dash pattern=on .7em off .3em}) to~$\column{f}_1^e$ or~$\mtrx{K}_{11}^e$, and dash-dot lines (\sampleline{dash pattern=on .7em off .2em on .05em off .2em}) to~$\mtrx{K}_{01}^e = (\mtrx{K}_{10}^e)^\mathsf{T}$ sub-matrices of~$\column{f}_\mathrm{M}^e$ or~$\mtrx{K}^e_\mathrm{M}$ in Eqs.~\eqref{eq:Mfint} and~\eqref{eq:Mk}; elements from Tab.~\ref{tab:integration} are used. Configuration of the element corresponds to homogeneous overall deformation~$\vec{\nabla}\vec{v}_0 = - 0.05\vec{e}_2\vec{e}_2$ with an affine micromorphic field~$v_1(\vec{X}) = (X_1+X_2+3)/2$, $X_i \in [-0.5,0.5]$, $i = 1,2$ (microstructure with a square stacking of holes used, i.e., $n=1$).}
	\label{fig:integrationdef}
\end{figure}

\begin{figure}
	\centering
	\begin{tabular}{@{}cccc@{}}
		\includegraphics[scale=0.7]{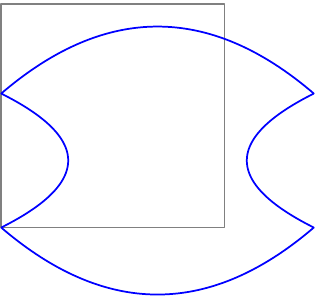} &
		\includegraphics[scale=0.7]{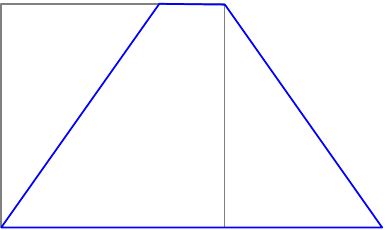} &
		\includegraphics[scale=0.7]{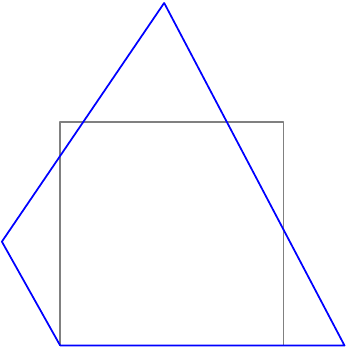} &
		\includegraphics[scale=0.9]{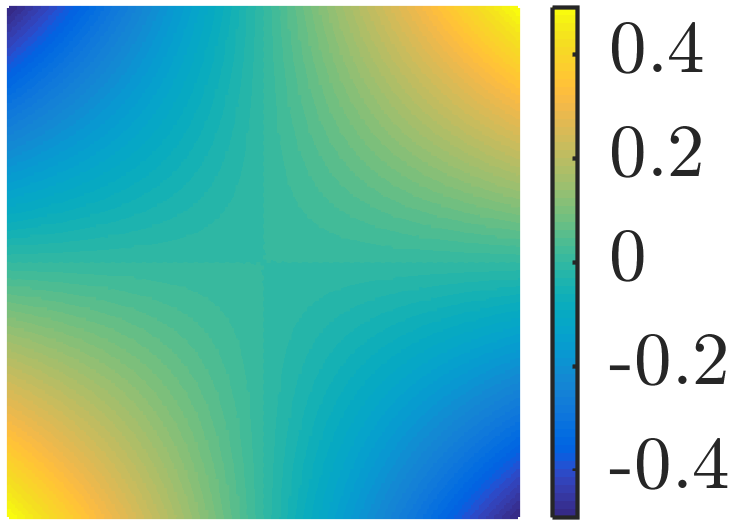} \\
		\tcaption{(a) Q8G4, $\vec{\psi}_1$} &
		\tcaption{(b) Q4G1, $\vec{\psi}_1$} &
		\tcaption{(c) Q4G1, $\vec{\psi}_2$} &
		\tcaption{(d) Q4G1, $\vec{\psi}_3(\vec{X})$}
	\end{tabular}
	\caption{A spurious singular mode observed for Q8G4 element in~(a), and three spurious singular modes for Q4G1 element in~(b)--(d) (typical hourglassing modes are shown also in Fig.~\ref{fig:intro}b). Figures (a)--(c) correspond to deformed configurations for which the micromorphic field vanishes, i.e.~$v_1(\vec{X}) = 0$, whereas~(d) shows a surface plot of~$v_1(\vec{X})$ for which the mean field vanishes, i.e.~$\vec{v}_0(\vec{X}) = \vec{0}$. Notation Q$i$G$j$ designates a quadrilateral element with~$i$ nodes and~$j$ integration points. In all cases, bottom left node is fixed in both directions, whereas the right bottom node is fixed in the vertical direction to eliminate rigid body modes. A microstructure with a square stacking of holes used, i.e., $n = 1$.}
	\label{fig:modes}
\end{figure}

To further support the above considerations and to extend their validity to the non-linear regime, we perform a numerical test on a metamaterial with a square stacking of holes, shown in Fig.~\ref{fig:intro}a and~\ref{fig:geometry}b. In such a case, only one patterning mode occurs (i.e.~$n = 1$), and the analytical expression of the mode~$\vec{\varphi}_1$ can be found e.g.~in~\cite[Eq.~(7)]{Rokos2018}. A single element in the reference configuration ($\vec{v}_0 = \vec{0}$, $v_1 = 0$) and in a deformed configuration ($\vec{\nabla}\vec{v}_0 = - 0.05\vec{e}_2\vec{e}_2$, $v_1(\vec{X}) = (X_1+X_2+3)/2$, $X_i \in [-0.5,0.5]$, $i = 1,2$) is considered. Errors in the internal forces (of Eq.~\eqref{eq:Mfint}) and stiffness matrix (of Eq.~\eqref{eq:Mk}) relative to those obtained with the highest order of integration (chosen as~$n_\mathrm{g} = 13$ for triangles and~$n_\mathrm{g} = 49$ for quadrilaterals) is shown in Fig.~\ref{fig:integrationref} for the reference and in Fig.~\ref{fig:integrationdef} for the deformed configuration as a function of the number of integration points. The results of Fig.~\ref{fig:integrationref} show that the error quickly drops to zero, proper integration rules coincide with those of Tab.~\ref{tab:integration}, and that the stiffness-equivalent integration rule provides error as low as~$0.5\,\%$ for triangles and~$5\,\%$ for quadrilaterals. Integration rules for which spurious singular modes occur are emphasized by red circles, and correspond to T6G1, and Q4G1, Q8G1, Q8G4, where the notation T$i$G$j$ (or Q$i$G$j$) stands for a triangular (or a quadrilateral) element with~$i$ nodes and~$j$ integration points. Spurious singular modes for Q4G1 and Q8G4 quadrilaterals are shown in Fig.~\ref{fig:modes}. The results of Fig.~\ref{fig:integrationdef} show a more complicated dependence, although observed errors drop quickly with increasing number of integration points for all tested element types.

The eight-node quadrilateral with four integration points, Q8G4, shows a non-communicable mode present only in the~$\vec{v}_0$ field. This mode is observed also in the standard FE formulations, but is usually not of any concern as it typically appears for a single element only and not in an element patch, cf. Fig.~\ref{fig:modes}a, \cite[Section~5.5.7.]{Bathe:2006}, and~\cite{Fan:1992}.
%
%
\subsection*{Q4G1 element}
\label{sect:Q4G1}
A question now arises whether the Q4G1 element could be stabilized to obtain an efficient element with a single Gauss integration point only for the micromorphic computational homogenization. As it may be seen from Figs.~\ref{fig:modes}b--\ref{fig:modes}d, all observed spurious singular modes correspond to the standard hourglassing mode, see Fig.~\ref{fig:intro}b and e.g.~\cite{Flanagan:1981}. In the context of the micromorphic homogenization, the hourglassing mode occurs twice in the mean field~$\vec{v}_0$, and once for each micromorphic field~$v_i$, thus inducing~$2+n$ rank deficiency of the resulting stiffness matrix in addition to proper deficiency of the three RBM.

Following standard stabilization procedures, see e.g.~\cite{Flanagan:1981,Belytschko:1983,Liu:1984,Belytschko:1984,Belytschko:1986,Hueck:1995}, stabilized micromorphic Q4G1 element stiffness matrix can be written as
\begin{equation}
\mtrx{K}^{e,\star}_\mathrm{M} = \mtrx{K}^e_\mathrm{M} + \mtrx{K}^\mathrm{stab},
\label{eq:stab}
\end{equation}
where~$\mtrx{K}^e_\mathrm{M}$ is obtained from Eq.~\eqref{eq:Mk} with~$n_\mathrm{g} = 1$. The stabilization matrix~$\mtrx{K}^\mathrm{stab}$ has the following form
\begin{equation}
\mtrx{K}^\mathrm{stab} = 
\begin{bmatrix}
\alpha_1\column{\gamma}\column{\gamma}^\mathsf{T} &  \alpha_2\column{\gamma}\column{\gamma}^\mathsf{T} & \mtrx{0} & \mtrx{0} & \dotsc & \mtrx{0} \\
\alpha_2\column{\gamma}\column{\gamma}^\mathsf{T} &  \alpha_1\column{\gamma}\column{\gamma}^\mathsf{T} & \mtrx{0} & \mtrx{0} & \dotsc & \mtrx{0} \\
\mtrx{0} & \mtrx{0} & \beta_1\column{\gamma}\column{\gamma}^\mathsf{T} & \mtrx{0} & \hdots & \mtrx{0} \\
\mtrx{0} & \mtrx{0} & \mtrx{0} & & & \\
\vdots & \vdots & \vdots & & \ddots & \vdots \\
\mtrx{0} & \mtrx{0} & \mtrx{0} & & \dotsc & \beta_n\column{\gamma}\column{\gamma}^\mathsf{T}
\end{bmatrix},
\label{eq:stabMatrix}
\end{equation}
where
\begin{align}
\column{\gamma} &= \frac{1}{4} \Big( \column{h} - (\column{h}^\mathsf{T}\column{x})\column{b}_x - (\column{h}^\mathsf{T}\column{y})\column{b}_y \Big), \label{eq:gamma}\\
\column{h} &= \begin{bmatrix} 1 & -1 & 1 & -1 \end{bmatrix}^\mathsf{T}, \label{eq:hourglass}
\end{align}
$\column{x}$, $\column{y}$ store nodal coordinates of a Q4 element in the reference configuration, $\column{b}_x$, $\column{b}_y$ are discrete versions of the gradient operator, and $\alpha_1$, $\alpha_2$, and~$\beta_i$, $i = 1,\dots,n$, are stabilization constants. An example of explicit stabilization values for standard continuum can be found e.g.~in~\cite[Tab.~1]{Belytschko:1986}. For simplicity, all mutual coupling terms between~$\vec{v}_0$ and~$v_i$ fields have been neglected, although a more in-depth derivation would yield off-diagonal stabilization terms as well. Such considerations are, however, outside the scope of this study and are left for future considerations.

In what follows, the formulation by Reese~\cite{Reese:2003} is adopted with small amendments. This formulation was originally developed for the stabilization of quadrilateral elements in large-deformation thermo-mechanically coupled problems, in which a scalar temperature field is considered in addition to a vector displacement field. A direct link between temperature and the micromorphic fields~$v_i$ can be thus established. The mechanical part of the stabilization matrix from~\eqref{eq:stabMatrix} reads
\begin{equation}
\mtrx{K}^\mathrm{stab}_{00} = 
\begin{bmatrix}
\alpha_1\column{\gamma}\column{\gamma}^\mathsf{T} & \alpha_2\column{\gamma}\column{\gamma}^\mathsf{T} \\
\alpha_2\column{\gamma}\column{\gamma}^\mathsf{T} & \alpha_1\column{\gamma}\column{\gamma}^\mathsf{T}
\end{bmatrix}
=
\mtrx{K}_{uu,\mathrm{hg}} - \mtrx{K}_{uv}\mtrx{K}_{vv}^{-1}\mtrx{K}_{vu},
\label{eq:mechStab}
\end{equation}
where the first term on the right-hand side~$\mtrx{K}_{uu,\mathrm{hg}}$ corresponds to hourglass stabilization, while the Schur complement accounts for the contribution of the incompatible (enhanced) assumed strain, not elaborated here in more detail. The constitutive stabilization tangent (denoted as~$\bs{A}_0^\star$ in~\cite{Reese:2003}) is taken as~$\overline{\mtrx{C}}$ (defined in Eq.~\eqref{eq:K00}), and updated only once at the beginning of the simulation. Although the stabilization matrix~$\mtrx{K}^\mathrm{stab}_{00}$ can be expressed by a pair of stabilization constants~$\alpha_1$ and~$\alpha_2$ \cite{Belytschko:1986}, we do not provide here their explicit forms, and rather define them implicitly through the second expression on the right hand side of Eq.~\eqref{eq:mechStab} due to the presence of the Schur complement term~$\mtrx{K}_{uv}\mtrx{K}_{vv}^{-1}\mtrx{K}_{vu}$. The structure of the micromorphic stabilization matrices reads
\begin{equation}
\mtrx{K}^\mathrm{stab}_{ii} = \beta_i\column{\gamma}\column{\gamma}^\mathsf{T} = 
\int_{\Omega_e} \underline{\gamma} \underline{L}_\mathrm{hg}^\mathsf{T} \mtrx{j}_0^\mathsf{T} \mtrx{E}_{ii}^\star \mtrx{j}_0 \underline{L}_\mathrm{hg} \underline{\gamma}^\mathsf{T} \,\mathrm{d}\vec{X},
\label{eq:micromorphichStab}
\end{equation}
where the stabilization coefficient~$\beta_i$ can be expressed in an explicit form as
\begin{equation}
\beta_i = 
\int_{\Omega_e} \underline{L}_\mathrm{hg}^\mathsf{T} \mtrx{j}_0^\mathsf{T} \mtrx{E}_{ii}^\star \mtrx{j}_0 \underline{L}_\mathrm{hg} \,\mathrm{d}\vec{X},
\label{eq:micromorphichStabBeta}
\end{equation}
and where~$\column{L}_\mathrm{hg} = [\eta,\xi,\eta,\xi]^\mathsf{T}$, with~$\xi,\,\eta \in [-1,1]$ being local coordinates on the parent element, $\mtrx{j}_0$ is a matrix storing entries of the Jacobian of the isoparametric mapping evaluated at the centre of the element, $\mtrx{E}_{ii}^\star$ is taken as~$\overline{\mtrx{E}}_{ii}$ (introduced in Eq.~\eqref{eq:Kii}, see also Eq.~\eqref{eq:Mstiffnessesc}) and again updated only once at the beginning of the simulation, whereas the integral is taken over an element's volume~$\Omega_e$ using the full four-point Gauss integration rule. For a thorough definition of all quantities involved in Eqs.~\eqref{eq:mechStab}--\eqref{eq:micromorphichStabBeta} and for further details we refer to~\cite{Reese:2003}.
%
%
\section*{Performance evaluation}
\label{sect:performance}
In this section, performance of individual elements is tested in terms of convergence properties and computational efficiency. A simple plane strain bending example shown in Fig.~\ref{fig:geometry} is considered, having a microstructure consisting of a square stacking of circular holes in an otherwise homogeneous hyperelastic matrix. Constitutive behaviour of the base material is specified by the following elastic energy density function
\begin{equation}
W(\tensor{F}) = a_1 (I_1 - 3) + a_2 (I_1 - 3)^2 - 2 a_1 \log{J} + \frac{1}{2} \kappa (J-1)^2,
\label{eq:constitutiveLaw}
\end{equation}
where~$\tensor{F} = \tensor{I} + (\vec{\nabla}\vec{u})^\mathsf{T}$ is the deformation gradient, $J = \det{\tensor{F}}$, $\tensor{I}$ is the second-order identity tensor, and~$I_1 = \tr{\tensor{C}}$ is the first invariant of the right Cauchy--Green deformation tensor~$\tensor{C} = \tensor{F}^\mathsf{T} \cdot \tensor{F}$. Constitutive parameters are chosen according to~\cite{Bertoldi2008d} as~$ a_1 = 0.55~$~MPa, $ a_2 = 0.3$~MPa, and~$\kappa = 55$~MPa. The adopted RVE is shown in Fig.~\ref{fig:geometry}b, having cell size~$\ell = 1$ and diameter~$d = 0.85\,\ell$, whereas the only patterning mode~$\vec{\varphi}_1$ (i.e.~$n = 1$) is shown in Fig.~\ref{fig:intro}a (for analytical expression we again refer to~\cite[Eq.~(7)]{Rokos2018}). Distributed external loading, applied at the right vertical edge of the cantilever specimen, was prescribed in the form
\begin{equation}
\vec{f}(\vec{X}) = -4t \left(1-\frac{4X_2^2}{H^2} \right)\vec{e}_2, \quad t \in [0,1],
\label{eq:loading}
\end{equation}
where~$t$ is parametrization time. Note that a similar problem was also considered in~\cite[Section~8.8.2]{BelytschkoFEM:2000}.

Four element types with various integration rules are used for the discretization of the macroscopic domain~$\Omega$, namely linear triangles T3G1 and T3G3, quadratic triangles T6G3 and T6G6, bi-linear quadrilaterals Q4G4 and the stabilized quadrilateral Q4G1, and serendipity quadratic quadrilaterals Q8G9 and Q8G4. For all element types, five element sizes~$h \in \{1, 2, 4, 8, 16\} \, \ell$ are used, discretized with right-angled triangles and rectangular quadrilaterals of good aspect ratios. For the discretization of the RVE problem, quadratic triangular elements of a typical size~$\ell/5$ with three Gauss integration points are used.
\begin{figure}
	\centering
	\begin{tabular}{@{}cc@{}}
		\begin{tikzpicture}
		\def\W{8} 
		\def\H{1.5} 
		\def\cellsize{0.82} 
		\def\diameter{0.867*\cellsize} 
		\coordinate (A) at (0,0);
		\coordinate (B) at (\cellsize*\W,0);
		\coordinate (C) at (\cellsize*\W,\cellsize*\H);
		\coordinate (D) at (0,\cellsize*\H);
		\filldraw[draw = white, fill = blue!5] (A) -- (B) -- (C) -- (D) -- cycle;
		\draw[thin] (D) -- (C);
		\draw[thin] (A) -- (B);
		\draw[thin] (A) -- (D);
		\draw[thin] (B) -- (C);
		\def\hcons{0.2}
		\draw[thin] (0,-0.25*\cellsize) -- (0,{\cellsize*(\H+0.2)});
		\foreach \i in {-0.25,0,...,\H}{%
			\draw[thin] (-0.2,{\cellsize*\i}) -- (0,{\cellsize*(\i+0.2)});}
		\draw[thin,latex'-latex'] (-1,0) -- (-1,\cellsize*\H) node[midway] {\footnotesize\setlength{\fboxsep}{3pt}\colorbox{white}{$H = 16\ell$}};
		\draw[thin] (-1.3,0) -- (-0.5,0);
		\draw[thin] (-1.3,\cellsize*\H) -- (-0.5,\cellsize*\H);
		\draw[thin,latex'-latex'] (0,-\cellsize*\H/2-0.2) -- (\cellsize*\W,-\cellsize*\H/2-0.2) node[midway,above] {\footnotesize $W = 128\ell$};
		\draw[thin] (0,{-\cellsize*\H/2-0.5}) -- (0,{-\cellsize*\H/2+0.3});
		\draw[thin] (\cellsize*\W,{-\cellsize*\H/2-0.5}) -- (\cellsize*\W,{-\cellsize*\H/2+0.3});
		\draw[-latex',line width=0.3mm] (0.5*\W*\cellsize,0.5*\H*\cellsize) -- (0.5*\W*\cellsize,0.5*\H*\cellsize+1) node[midway,anchor =  east,shift={(0,0.4)}] {\footnotesize $\vec{e}_{2}$};
		\draw[-latex',line width=0.3mm] (0.5*\W*\cellsize,0.5*\H*\cellsize) -- (0.5*\W*\cellsize+1,0.5*\H*\cellsize) node[midway,anchor = south,shift={(0.4,0)}] {\footnotesize $\vec{e}_{1}$};
		\draw[thin] (\W*\cellsize,\H*\cellsize) to [bend left=60] node[at start,right,anchor = south] {\footnotesize $\vec{f}(\vec{X})$} (\W*\cellsize,0);
		\def\arrowlength{0.25}
		\foreach \i in {0.25,0.5,0.75}{%
			\draw[thin,-latex'] (\W*\cellsize+0.1,{\cellsize*\i*\H+\arrowlength/2}) -- (\W*\cellsize+0.1,{\cellsize*\i*\H-\arrowlength/2});}
		\end{tikzpicture} &
		\begin{tikzpicture}
		\def\W{2} 
		\def\H{2} 
		\def\cellsize{1.2} 
		\def\diameter{0.85*\cellsize} 
		\coordinate (A) at (0,0);
		\coordinate (B) at (\cellsize*\W,0);
		\coordinate (C) at (\cellsize*\W,\cellsize*\H);
		\coordinate (D) at (0,\cellsize*\H);
		\filldraw[draw = white, fill = blue!5] (A) -- (B) -- (C) -- (D) -- cycle;
		\draw[thin] (D) -- (C);
		\draw[thin] (A) -- (B);
		\draw[thin] (A) -- (D);
		\draw[thin] (B) -- (C);
		\def\skippbc{0.05}
		\draw[thick,mygreen,dash pattern=on 6pt off 3pt] (\cellsize*\W+\skippbc,0) -- (\cellsize*\W+\skippbc,\cellsize*\H+\skippbc) -- (0,\cellsize*\H+\skippbc) node[midway,above] {\footnotesize $\partial\Omega_\mathrm{m}^+$};
		\draw[thick,myblue,dash pattern=on 6pt off 3pt] (\cellsize*\W+\skippbc,-\skippbc) -- (-\skippbc,-\skippbc) node[midway,below] {\footnotesize $\partial\Omega_\mathrm{m}^-$} --  (-\skippbc,\cellsize*\H+\skippbc);	
		\foreach \i in {1,...,\W}{%
			\foreach \j in {1,...,\H}{%
				\filldraw[fill=white,thin] ({\cellsize*(\i-0.5)},{\cellsize*(\j-0.5)}) circle ({\diameter/2});
		}}
		\draw[thin,latex'-latex'] ({\cellsize*(\W+0.25)},\cellsize) -- ({\cellsize*(\W+0.25)},2*\cellsize) node[midway,right] {\footnotesize $\ell$};
		\draw[dashed] ({\cellsize*(\W+0.25)},\cellsize) -- ({\cellsize*(\W-1)},\cellsize) -- ({\cellsize*(\W-1)},2*\cellsize) -- ({\cellsize*(\W+0.25)},2*\cellsize);
		\draw[thin,latex'-latex'] ({\cellsize*(\W-0.5)},{\cellsize*(1+0.0665)}) -- ({\cellsize*(\W-0.5)},{\cellsize*(2-0.0665)}) node[midway,right,shift={(-0.1,0)}] {\footnotesize $d$};
		\end{tikzpicture} \\
		\tcaption{(a) cantilever beam problem} &
		\tcaption{(b) considered RVE}
	\end{tabular}
	\caption{(a)~Considered benchmark example of a cantilever beam subjected to bending. (b)~Representative volume element associated with each macroscopic Gauss integration point (the corresponding patterning mode is shown in Fig.~\ref{fig:intro}a).}
	\label{fig:geometry}
\end{figure}
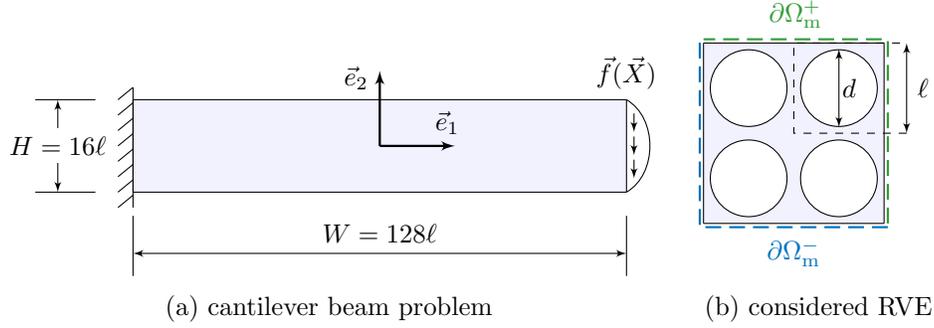
\begin{figure}
	\centering
	\begin{tabular}{@{} m{6cm} @{} m{6cm} @{}}
		\begin{tikzpicture}
		\node[inner sep=0,outer sep=0] (figure) {\includegraphics[scale=0.9]{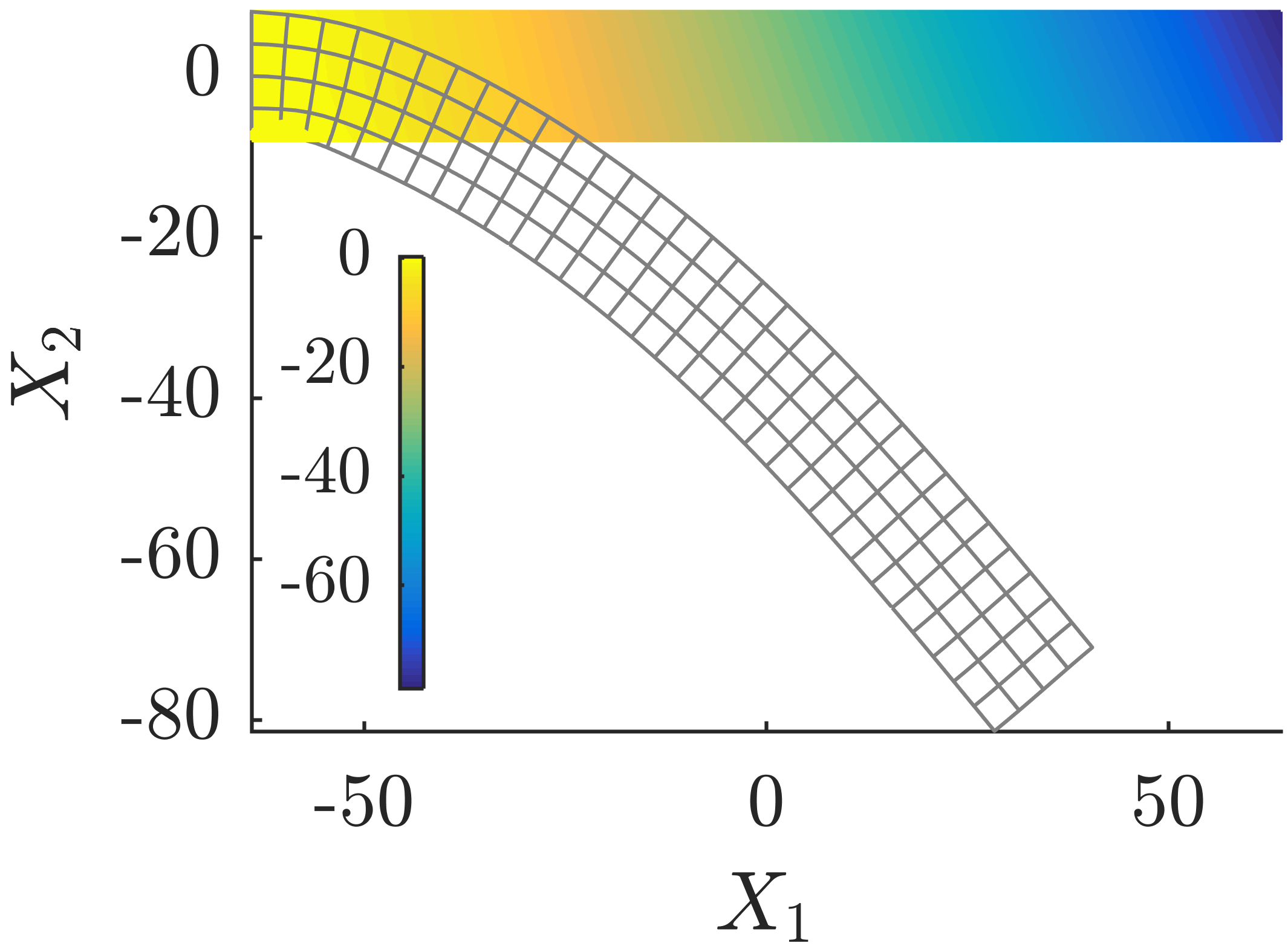}};
		\node[xshift=-0.4cm,yshift=-0.8cm] (label) {\footnotesize $\vec{v}_0\cdot\vec{e}_2$};
		\end{tikzpicture} & 
		\begin{tikzpicture}
		\node[inner sep=0,outer sep=0] (figure) {\includegraphics[scale=0.9]{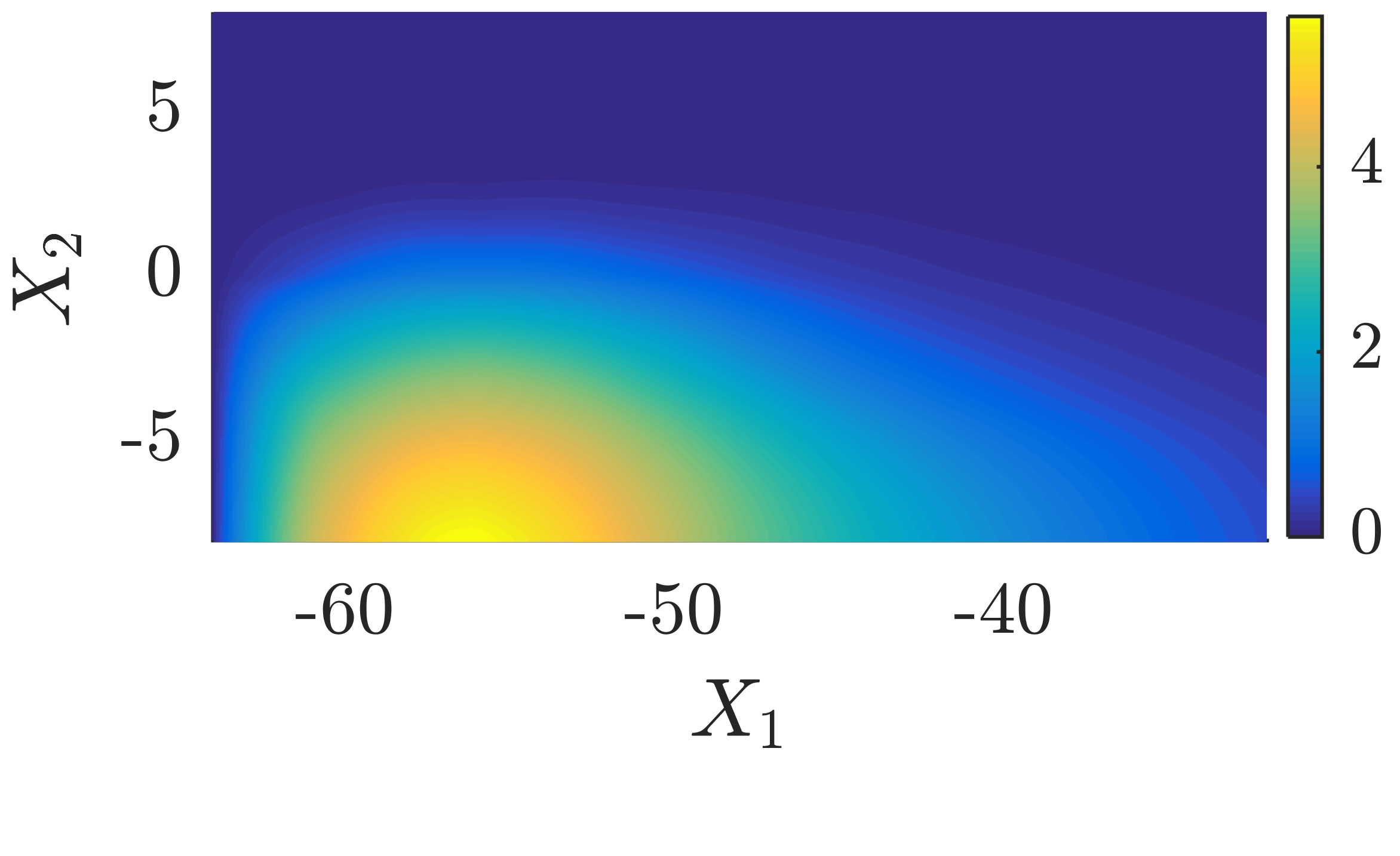}};
		\node[xshift=2.65cm,yshift=-0.8cm] (label) {\footnotesize $v_1$};
		\end{tikzpicture} \\
		\centering\tcaption{(a) deformed configuration, $t = 1$} &
		\centering\tcaption{(b) $v_1(\vec{X})$, $t = 1$, a close-up}
	\end{tabular}	
	\caption{(a)~Deformed mesh of~$\vec{v}_0(\vec{X})$; colour plot shows~$\vec{v}_0\cdot\vec{e}_2$. (b) Corresponding field plot of~$v_1(\vec{X})$, close-up on the left quarter of the specimen. In both cases, $t = 1$, and the third finest discretization (i.e.~$h = 4\ell$) using Q8G9 elements for the bending example shown in Fig.~\ref{fig:geometry}a is used.}
	\label{fig:deformed}
\end{figure}

The numerical solution of the macroscopic problem obtained for Q8G9 elements with~$h = 4\ell$ at~$t = 1$ appears in Fig.~\ref{fig:deformed}. From the micromorphic field~$v_1$ shown in Fig.~\ref{fig:deformed}b we can clearly see that the patterning mode is triggered only in the compressive region at the bottom-left part of the specimen domain, and decays rapidly in the close vicinity of prescribed displacements (the left vertical edge). Although the obtained results correspond to expectations, no comparison with a direct numerical simulation fully accounting for the underlying microstructure is available. Instead, the solution corresponding to the T6G6 element with~$h = 1\ell$ is considered as the reference solution against which the accuracy of other discretizations is measured.

\begin{figure}
	\centering
	\includegraphics[scale=0.9]{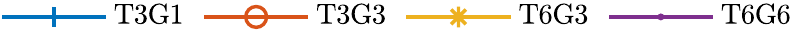}\\
	\includegraphics[scale=0.9]{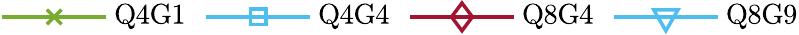}\\	
	\vspace{0.5em}
	\begin{tabular}{@{}cc@{}}
		\includegraphics[scale=0.9]{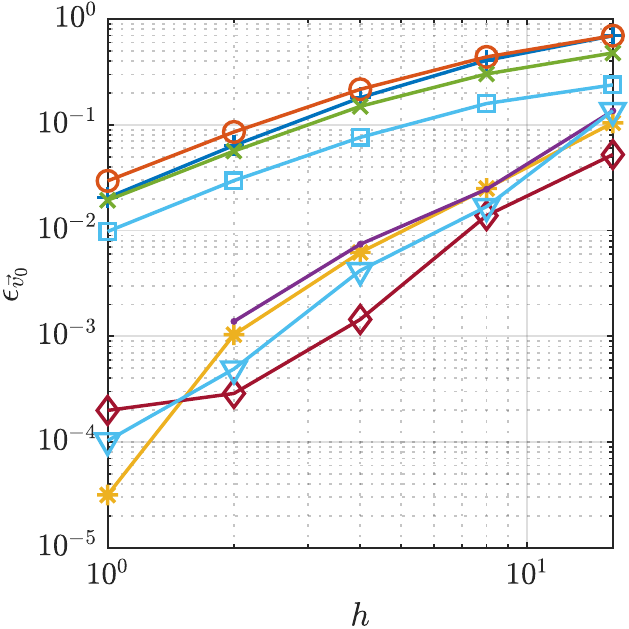} & 
		\includegraphics[scale=0.9]{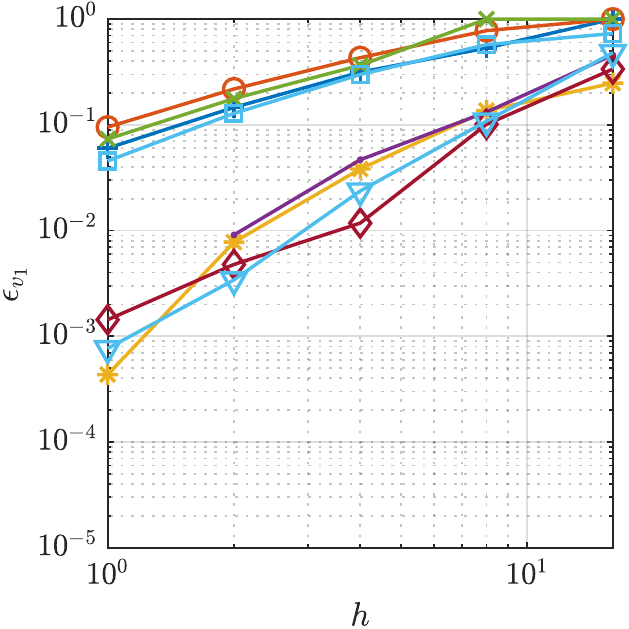} \\
		\tcaption{(a) error in~$\vec{v}_0$} &
		\tcaption{(b) error in~$v_1$} \\
		\includegraphics[scale=0.9]{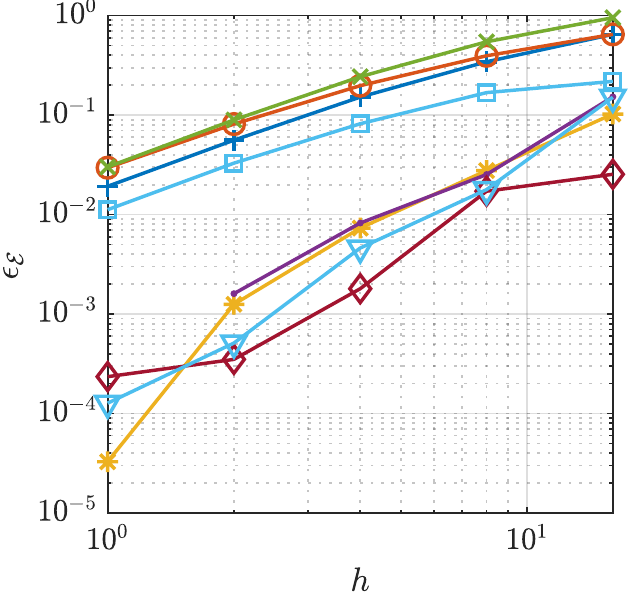} & 
		\includegraphics[scale=0.9]{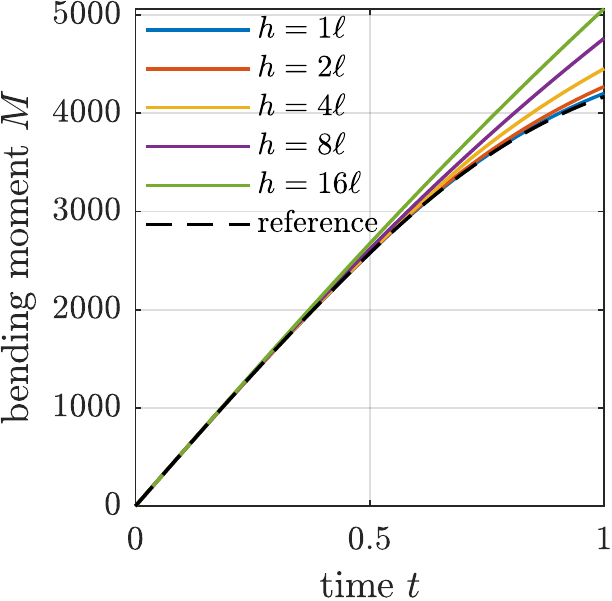} \\
		\tcaption{(c) error in energy~$\mathcal{E}$} &
		\tcaption{(d) Q4G1, reaction bending moment} \\	
	\end{tabular}
	\caption{Convergence of~(a) $L_2$ displacement error in~$\vec{v}_0(\vec{X})$, $t = 1$, (b) $L_2$ micromorphic error in~$v_1(\vec{X})$, $t = 1$, and~(c) error in energy evolutions~$\mathcal{E}(t)$, $t \in [0,1]$. (d)~Reaction bending moment~$M(t)$ for Q4G1 and various element sizes~$h$ in comparison with the reference solution. Data points for T6G6, $h=1\ell$, are missing, because this solution is considered as the reference solution and the corresponding error thus vanishes.}
	\label{fig:convergence}
\end{figure}

Fig.~\ref{fig:convergence} shows convergence plots of the relative error for individual element types, $\epsilon$, which is defined as
\begin{equation}
\epsilon_g = \frac{\| g - g_\mathrm{ref} \|_2}{\| g_\mathrm{ref} \|_2},
\label{eq:error}
\end{equation}
where~$g$ denotes a quantity of interest ($\vec{v}_0$, $v_1$, or~$\mathcal{E}$), and~$g_\mathrm{ref}$ is its corresponding reference value represented by the T6G6, $h = 1\ell$, solution. In particular, Fig.~\ref{fig:convergence}a shows error in the mean field~$\vec{v}_0$ as a function of the element size~$h$ evaluated at the end of the loading history, i.e.~at~$t = 1$. Similarly, Fig.~\ref{fig:convergence}b shows the error evaluated for the micromorphic field, whereas Fig.~\ref{fig:convergence}c captures the error in energy evolutions evaluated as a function of time. Observed trends for all three figures are very similar: the linear and bi-linear elements (T3, Q4) converge slower than the quadratic elements (T6, Q8), ordered according to their level of interpolation.

\begin{table}
	\begin{minipage}[t]{\linewidth}
		\let\footnoterule\relax
		\centering
		\caption{Hourglass energy~$\mathcal{H}$ relative to the total elastic energy~$\mathcal{E}$ of the specimen for Q4G1 element as a function of the element size~$h \in \{1, 2, 4, 8, 16\} \, \ell$.}
		\renewcommand*{\arraystretch}{1.3}
		\label{tab:hourglassEnergy}
		\begin{tabular}{cccccc}\hline
			$h$ & 16$\ell$ & 8$\ell$ & 4$\ell$ & 2$\ell$ & 1$\ell$ \\\hline
			$\mathcal{H}/\mathcal{E}$ & 21.8 & 1.98 & 0.304 & 0.0960 & 0.0138 \\\hline
		\end{tabular}
	\end{minipage}
\end{table}
\begin{table}
	\begin{minipage}[t]{\linewidth}
		\let\footnoterule\relax
		\centering
		\caption{Computing times~$T$ relative to the fastest discretization (Q4G1) for~$h = 2\ell$, number of macroscopic elements~$n_e$, Gauss integration points~$n_\mathrm{g}$, and nodes~$n_\mathrm{node}$.}
		\renewcommand*{\arraystretch}{1.3}
		\label{tab:efficiency}
		\begin{tabular}{ccccccccc}\hline
			el. type & T3G1 & T3G3 & T6G3 & T6G6 & Q4G1 & Q4G4 & Q8G4 & Q8G9 \\\hline
			$T$ & 1.9 & 5.6 & 6.1 & 11.6 & 1 & 3.7 & 4.4 & 9.4 \\
			$n_e$ & 1024 & 1024 & 1024 & 1024 & 512 & 512 & 512 & 512 \\
			$n_\mathrm{g}$ & 1024 & 3072 & 3072 & 6144 & 512 & 2048 & 2048 & 4608 \\			
			$n_\mathrm{node}$ & 585 & 585 & 2193 & 2193 & 585 & 585 & 1681 & 1681 \\\hline
		\end{tabular}
	\end{minipage}
\end{table}
\begin{figure}
	\centering
	\begin{tabular}{@{} m{5.41cm} @{} m{7cm} @{}}
		\begin{tikzpicture}
		\node[inner sep=0,outer sep=0] (figure) {\includegraphics[scale=0.9]{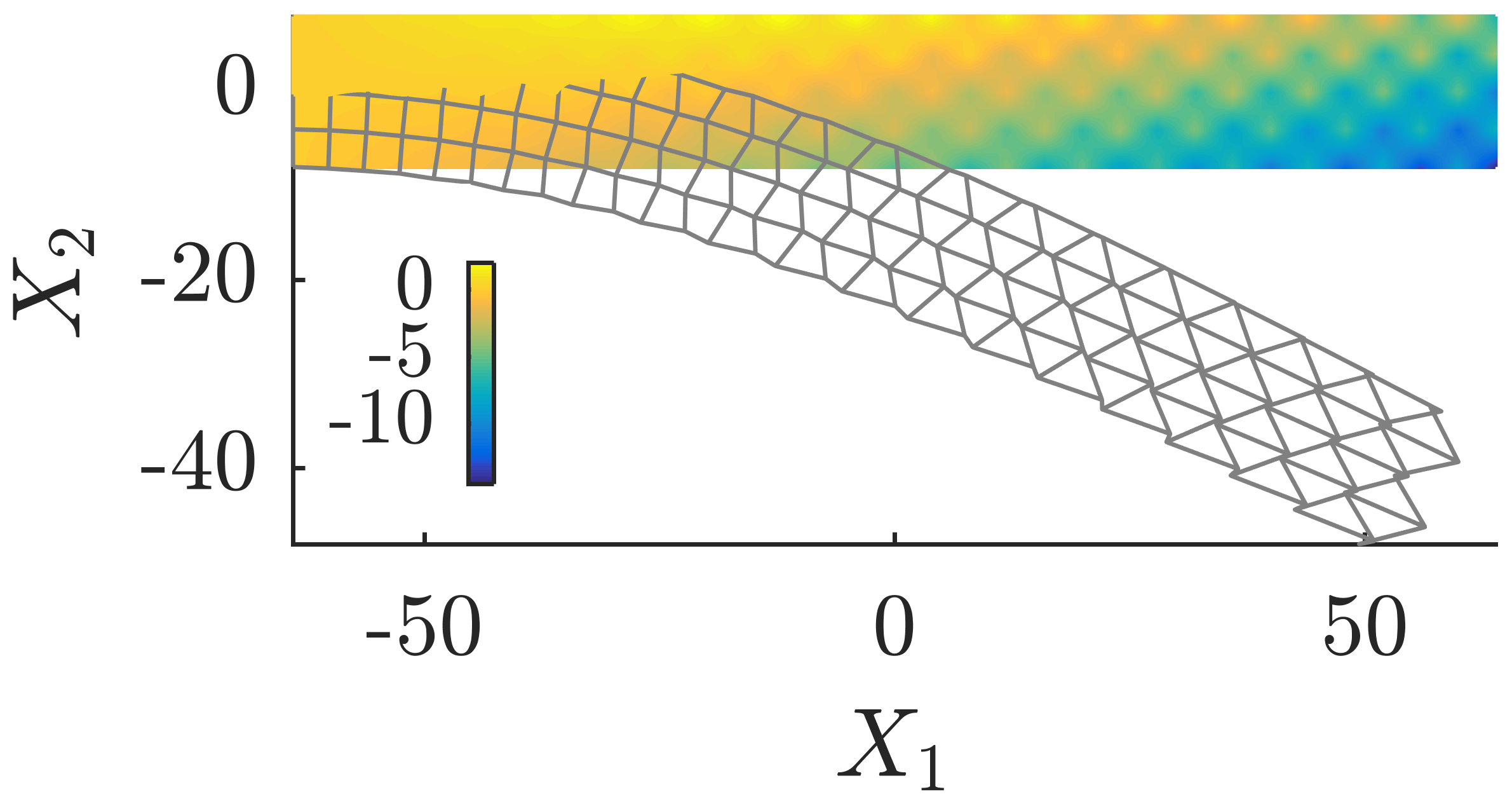}};
		\node[xshift=-0.4cm,yshift=-0.2cm] (label) {\footnotesize $\vec{v}_0\cdot\vec{e}_1$};
		\end{tikzpicture} & 
		\begin{tikzpicture}
		\node[inner sep=2pt,outer sep=0] (figure) {\includegraphics[scale=0.9]{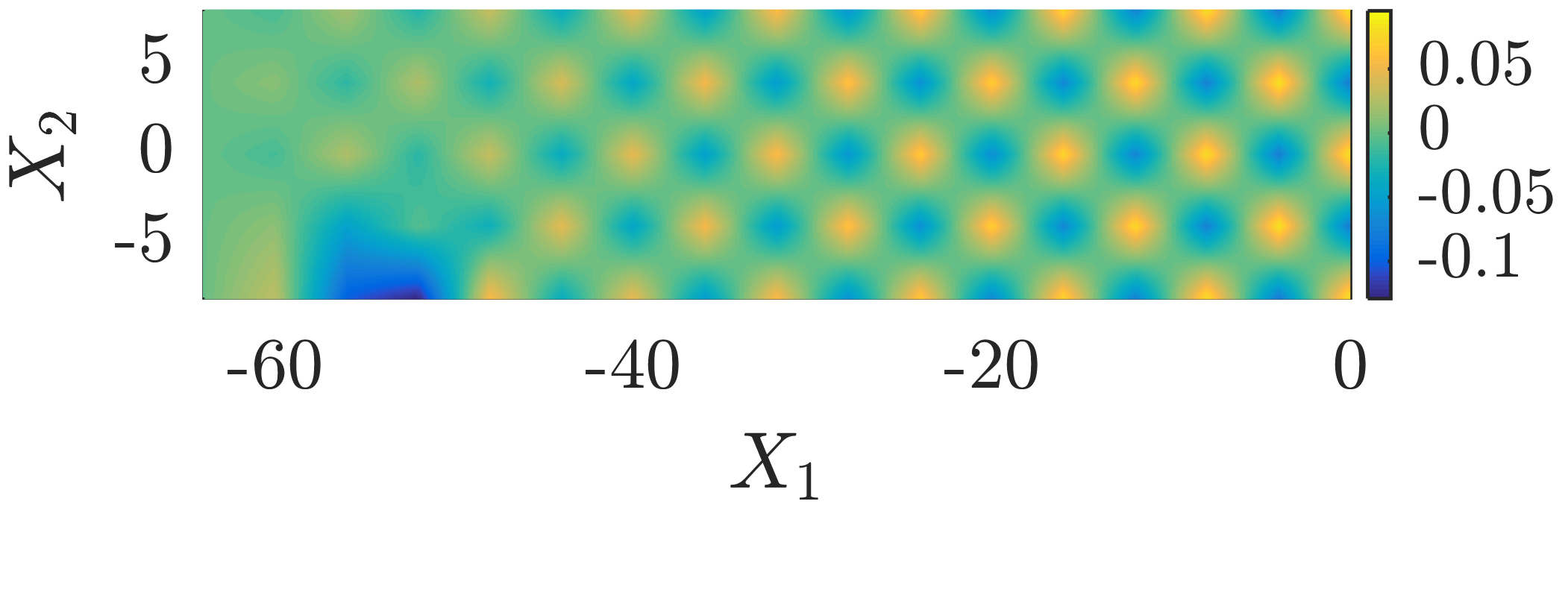}};
		\node[xshift=2.9cm,yshift=-0.2cm] (label) {\footnotesize $v_1$};
		\end{tikzpicture} \\
		\centering\tcaption{(a) deformed configuration, $t = 0.5$} &
		\centering\tcaption{(b) buckling mode, $v_1(\vec{X})$, close-up}
	\end{tabular}	
	\caption{Example of wrong choices of stabilization parameters for Q4G1, $h = 4\ell$. (a)~Artificial instability polluting the mean field~$\vec{v}_0$; deformed configuration at~$t = 0.5$ obtained for~$\alpha_1 = \alpha_2 = 0$, where colour plot shows~$\vec{v}_0\cdot\vec{e}_1$. (b)~Artificial instability in the micromorphic field polluting a buckling mode obtained for~$\beta_1 = 0$. Recall the structure of~$\mtrx{K}^\mathrm{stab}$ in Eq.~\eqref{eq:stabMatrix}.}
	\label{fig:wrongPenalty}
\end{figure}

The under-integrated element Q4G1 reaches a similar rate of convergence as T3G1, T3G3, and Q4G4, although it is slightly less accurate, especially in the energy norm. Its corresponding convergence in the reaction moment is shown in Fig.~\ref{fig:convergence}d, where a good agreement for the two finest approximations~$h = 1\ell$ and~$h = 2\ell$ is observed. The hourglass energy required for stabilization of all elements, $\mathcal{H}$, relative to the elastic energy of the entire system, $\mathcal{E}$, is summarized in Tab.~\ref{tab:hourglassEnergy}. The entries presented suggest that due to the bending character of the deformation and loading, the stabilization energy reaches considerable values for coarse discretizations, being acceptable only for as fine discretizations as~$h = 2\ell$. 

Computing times~$T$ relative to the fastest discretization (Q4G1) as well as the number of macroscopic elements~$n_e$, Gauss integration points~$n_\mathrm{g}$, and nodes~$n_\mathrm{node}$, are summarized for~$h = 2\ell$ in Tab.~\ref{tab:efficiency}. Here we conclude that although Q4G1 achieves comparable accuracy as T3G1 and Q4G4, it is significantly faster. Another interesting combination is the element Q8G4, which is roughly four times slower in comparison with Q4G1, but achieves a significantly higher accuracy (recall Fig.~\ref{fig:convergence}).

Let us note finally that unlike the definitions of Eqs.~\eqref{eq:mechStab}--\eqref{eq:micromorphichStabBeta} which perform optimally with respect to incompatible enhanced strains, the stabilization factors~$\alpha_1$, $\alpha_2$, and~$\beta_i$, $i = 1, \dots, n$, can also be chosen manually. When their values are chosen too small or when they vanish altogether for~$\vec{v}_0$ (i.e.~$\alpha_1 = \alpha_2 = 0$), spurious instabilities pollute the mean solution. For comparison, such a situation is depicted in Fig.~\ref{fig:wrongPenalty}a, where a deformed configuration is shown along with the horizontal component of the mean field~$\vec{v}_0$ (in colour), featuring clear hourglass oscillations. When such oscillations become too severe, the  Newton solver even fails to converge. Upon choosing too weak or vanishing stabilization in the micromorphic field~$v_1$ while stabilizing the mean solution~$\vec{v}_0$ (i.e.~$\beta_1 = 0$), a buckling mode, needed to initialize the Newton solver towards a proper equilibrium path upon physical instability, is also distorted by the spurious hourglassing mode clearly seen in Fig.~\ref{fig:wrongPenalty}b. Spurious oscillations may occur even despite the proper stabilization, i.e. for sufficiently large~$\alpha_1$, $\alpha_2$, and~$\beta_i$; it is then recommended to switch (locally or globally) to the full four-point integration scheme.
%
%
\section*{Summary and conclusions}
\label{summary}
In this paper, proper choices of elements and integration rules for the micromorphic computational homogenization scheme have been discussed. It has been shown that although the micromorphic part of the stiffness matrix necessitates a mass-matrix equivalent integration rule, this requirement can be relaxed with only a negligible drop in accuracy and no artificial spurious modes. An efficient one-integration point quadrilateral element was further discussed, which offers the advantage of being computationally efficient, while maintaining a reasonable accuracy and convergence. Because standard hourglassing modes are observed for this type of element based on the eigenvalue analysis of the stiffness matrix, a suitable stabilization technique was introduced and discussed, based on methods available in the literature. A~dedicated stabilization technique might be developed to further improve the performance of the under-integrated element, which lies, however, outside of the scope of the current study. Using a benchmark numerical example it was further shown that the achieved performance of the one-integration point element is satisfactory in comparison to the full integration rule, requiring, nevertheless, approximately only one fourth of the computational effort in comparison to the fully integrated four-node quadrilateral. Another interesting option for the micromorphic computational homogenization scheme is the eight-noded quadrilateral with four integration points, which achieves good performance and high accuracy.
%
%
\section*{Abbreviations}
RVE: representative volume element; FE: finite element; FEM: finite element method; DOF: degree of freedom; RBM: rigid body modes; T$i$G$j$ (or Q$i$G$j$): a triangular (or a quadrilateral) element with~$i$ nodes and~$j$ integration points;
%
%
\section*{Author's contributions}
OR carried out most of the study, performed numerical simulations, and drafted the manuscript. MD helped with implementation and numerical issues. All authors developed the methodology, conceived of the study, and participated in its design, coordination, and critical review of the manuscript. All authors also read and approved the final manuscript.
%
%
\section*{Acknowledgements}
The research leading to these results has received funding from the Czech Science Foundation (GA\v{C}R) under grant agreement no.~[19-26143X]. PK would like to thank the Ministry of Education, Youth and Sports of the Czech Republic for financial support of his stay at the Czech Technical University in Prague in the framework of International Mobility of Researchers, project No.~CZ.02.2.69/0.0/0.0/16\_027/0008465.
%
%
%
\bibliography{mybibfile}
\end{document}